\documentclass[useAMS,usenatbib]{mn2e}
\input{psfig}
\usepackage{times}
\usepackage{enumitem}
\usepackage{bm}
\usepackage{amsmath} 
\usepackage{url}
\usepackage{amssymb}
\usepackage{dblfnote}
\usepackage{aas_macros}
\usepackage{hyperref}
\usepackage{natbib}
\usepackage{ifthen}
\def\Real{{\rm I\mathchoice{\kern-0.70mm}{\kern-0.70mm}{\kern-0.65mm}%
 {\kern-0.50mm}R}}

\font \bolditalics = cmmib10
\def\bx#1{\leavevmode\thinspace\hbox{vrule\vtop{\vbox{\hrule\kern1pt
 \hbox{\vphantom{\tt/}\thinspace{\bf#1}\thinspace}}
 \kern1pt\hrule}\vrule}\thinspace}

\def \vc #1{{\textfont1=\bolditalics \hbox{$\bf#1$}}}

\newcommand{\bea}{\begin{eqnarray}}
\newcommand{\eea}{\end{eqnarray}}
\newcommand{\be}{\begin{equation}}
\newcommand{\ee}{\end{equation}}

\def\bx{{\bf x}}
\def\bk{{\bf k}}

\def\thetag{{\vc \theta}}

\def\be{\begin{equation}}
\def\ee{\end{equation}}

\def\ba{\begin{eqnarray}}
\def\ea{\end{eqnarray}}

\def \vc #1{{\textfont1=\bolditalics \hbox{$\bf#1$}}}{\catcode`\@=11

\def\ave#1{\left\langle #1 \right\rangle}

\def\ltsima{$\; \buildrel < \over \sim \;$}
\def\lsim{\lower.5ex\hbox{\ltsima}}
\def\gtsima{$\; \buildrel > \over \sim \;$}
\def\gsim{\lower.5ex\hbox{\gtsima}}

\title[Shear estimation bias by colour gradients]{On the shear estimation bias induced by the spatial variation of colour across  galaxy profiles.}
\author[Semboloni et al.]{E. Semboloni$^{1}$\thanks{sembolon@strw.leidenuniv.nl}, H. Hoekstra$^{1}$, Z. Huang$^{2}$,  V.~F. Cardone$^{2}$, M. Cropper$^{3}$,
\newauthor{B. Joachimi$^{4}$, T. Kitching$^{3,4}$, K. Kuijken$^{1}$,   M. Lombardi$^{5}$, R. Maoli$^{6}$, Y. Mellier$^{7}$,}
\newauthor{L. Miller$^{8}$,  J. Rhodes$^9$, R. Scaramella$^{2}$, T. Schrabback$^{10,11}$,  M. Velander$^{1,8}$} \\
$^1$Leiden Observatory, Leiden University, P.O. Box 9513, 2300 RA, Leiden, The Netherlands \\
$^2$ INAF, Osservatorio Astronomico di Roma,  via Frascati 33, 00040, Monteporzio Catone, Italy  \\
$^3$  Mullard Space Laboratory, University College London, Holmbury St Mary, Dorking, Surrey RH5 6NT, UK  \\
$^4$ University of Edinburgh, Royal Observatory, Blackford Hill, Edinburgh EH9 3HJ, UK   \\
$^5$ Dipartimento di Fisica, Universit\`a degli Studi di Milano, via Celoria, 16, I-20133 Milano, Italy\\
$^6$ Dipartimento di Fisica, Universit\`a di Roma ``La Sapienza'', Piazzale AldoMoro 2, I-00185 - Roma, Italy\\
$^7$ Institut d'Astrophysique de Paris, UMR7095  CNRS, Universit\'e Pierre et Marie Curie, 98 bis Boulevard Arago, 75014 Paris, France \\
$^8$  Department of Physics, University of Oxford, The Denys Wilkinson Building, Keble Road, Oxford, OX1 3RH, UK  \\
$^9$  Jet Propulsor Laboratory, California Institute of Technology, 4800 Oak Grove Drive, Pasadena, CA 91109, USA \\
$^{10}$ Argelander-Institut f\"ur Astronomie, Auf dem H\"ugel 71, D-53121 Bonn, Germany \\
$^{11}$ Kavli Institute for Particle Astrophysics and Cosmology, Stanford
University, 382 Via Pueblo Mall, Stanford, CA 94305-4060, USA\\
}

\begin{document}
\maketitle
\begin{abstract}
The spatial variation of the colour of a galaxy may introduce a bias
in the measurement of its shape if the PSF profile depends on
wavelength.  We study how this bias depends on the properties of the
PSF and the galaxies themselves. The bias depends on the scales used
to estimate the shape, which may be used to optimise methods to reduce
the bias. Here we develop a general approach to quantify the
bias. Although applicable to any weak lensing survey, we focus on the
implications for the ESA {\it Euclid} mission. 

Based on our study of synthetic galaxies we find that the bias is a
few times $10^{-3}$ for a typical galaxy observed by {\it Euclid}.
Consequently, it cannot be neglected and needs to be accounted for.  We
demonstrate how one can do so using spatially resolved observations of galaxies
in two filters. We show that {\it HST} observations in the F606W and
F814W filters allow us to model and reduce the bias by an order of
magnitude, sufficient to meet {\it Euclid}'s scientific
requirements. The precision of the correction is ultimately determined
by the number of galaxies for which spatially-resolved observations in
at least two filters are available. We use results from the Millennium
Simulation to demonstrate that archival {\it HST} data will be
sufficient for the tomographic cosmic shear analysis with the {\it
  Euclid} dataset.

\end{abstract}


\section{Introduction}

The measurement of the  distortion of the shapes of galaxies caused by gravitational lensing  by large scale structures, i.e. cosmic
shear, is a powerful tool to investigate the statistical properties of
the Universe and in particular to understand the mechanism responsible
for the observed accelerated expansion.  The origin of this
acceleration has been dubbed ``dark energy'' reflecting the fact that
we still lack a theoretical framework to explain its nature.  To
discriminate between competing theories, large cosmological experiments are needed.
The most ambitious among these is the recently selected ESA mission,
{\it Euclid}\footnote{\url{http://www.euclid-ec.org}} \citep{redbook},
which will survey the 15,000 deg$^2$ of the extragalactic sky that has
both low extinction and zodiacal light.

The measurement of the second-order shear statistics as a function of
redshift, which is commonly referred to as cosmic shear tomography, is
particularly powerful to study the growth of structures and the
expansion history of the Universe (\citealt{Hu99,Hu02}; see
\citealt{HoJa08,Munshi08}, for recent reviews).  The accuracy of the
derived constraints on the cosmological parameters depends critically
on our ability to measure both the shapes and redshifts of the
billions of galaxies that will be observed. Both are challenging tasks
and {\it Euclid} has been designed to achieve this. For a detailed
discussion of how this can be done, we refer the reader to
\cite{Cropper12}.

A critical step in the estimation of the cosmic shear signal from the
observed shapes of galaxies is the correction for the point spread
function (PSF): as shown in \cite{Masseyetal12} most residual biases
scale proportional to the square of the PSF size. The biases affecting
shear measurement techniques have been studied in a number of
collaborative projects such as the Shear TEsting Programme
\citep[STEP;][]{Heymansetal06,Masseyetal07}, the Gravitational LEnsing
Accuracy Testing 2008 \citep[GREAT08;][]{Bridleetal09,Bridleetal10}
and the GREAT10 challenges
\citep{Kitchingetal10,Kitchingetal12a,Kitchingetal12b} using simulated
monochromatic data. The impact of the noise in the data, which affects
the fainter galaxies, has been recently studied
\citep{Refregieretal12,Kaetal12,MeVi12}. While these studies are
necessary to calibrate methods so that they will perform increasingly
well on monochromatic images, we explore here another source of bias
which has not been considered by those studies.

Cosmic shear surveys are designed to maximise the number of observed
galaxies. For this reason they use broad-band filters; the observed
images are the integrated light distribution over a large wavelength
range. The spectral energy distribution (SED) of a galaxy typically
varies spatially, generating ``colour gradients''. This  prevents one from unambiguously recovering the unconvolved light distribution required for an unbiased shear estimate from the observed images.

The existence of colour gradients is one potential source
of systematic error diminishing the power of future weak lensing
missions.  Observations of galaxies in a broad-band filter to estimate
the shear add some complications even if galaxies have no colour
gradients. This happens because the PSF is chromatic. Ignoring this fact leads to a bias which was quantified by
\cite{Cyprianoetal10}. This bias can be avoided  using a colour weighted  PSF, that   depends on the global SED
of the galaxy.  In the case of {\it Euclid} this PSF can be estimated with
sufficient accuracy from supporting deep multi-colour ground-based
observations.

While the PSF correction described by \cite{Cyprianoetal10} is perfect
when the galaxies have no colour gradients, it becomes inaccurate when
the colour varies spatially. This can be understood by considering the
simple case of a galaxy with a small red bulge and a much more
extended blue disk. The global colour that is used to estimate the
SED-weighted PSF will resemble that of the disk if its flux exceeds
that of the bulge. To suppress noise, shape measurements employ a
radial weighting that is matched to the brightness profile of the
galaxy. This enhances the contribution of the bulge to the shape
measurement, implying that a PSF redder than the SED-weighted one
should be used.

As we will quantify in detail below, the bias depends on the width of
the filter that is used. Consequently, colour gradients are expected to
be particularly relevant for {\it Euclid} because of its wide pass-band
\citep{redbook}. We note, however, that it may not even be negligible
for future ground-based experiments. Despite the fact that
multi-wavelength ground-based projects will employ narrower filters
and that the wavelength dependence of the PSF is weaker, the
intrinsically larger PSF exacerbates the impact of colour gradients
\citep[see][for a detailed analysis of the impact of the PSF size and
  other systematics on shape measurements]{Masseyetal12}.

The impact of colour gradients on shear measurements was first studied
by \cite{Voigtetal12} who used a small sample of {\it HST} galaxies
for which a bulge plus disk decomposition was available
\citep{Simardetal02}. \cite{Voigtetal12} 
quantified the bias induced by colour gradients and also
estimated the minimum size of a sample of galaxies that needs to be
observed in two narrower filters in order to determine the bias.
Because of the small sample size and a concern that the  bulge plus
disk decomposition might not fully capture the properties of real
galaxies, \cite{Voigtetal12} provided conservative upper limits.

The aim of this paper is to expand upon this pioneering study. To this end
we develop an approach that allows us to explore how the bias depends
on the wavelength range covered by the filter and on the properties of
the PSF and the galaxies. Furthermore, we discuss how the amplitude of
the bias depends on the method used to estimate the shear. To do so,
we use simulated galaxies and show how the bias can be modelled. Our
methodology can in principle be applied to archival {\it HST} data or any
other dataset of well-resolved galaxies observed in least two filters.
We argue that the {\it HST} archive represents the most suitable
dataset for colour gradient studies because of its PSF
characteristics. We show that a sufficient number of
galaxies has been observed to calibrate the colour gradient induced
bias with the precision required to achieve {\it Euclid}'s science
objectives.

The paper is organised as follows: in Section \ref{sec:intro} we
introduce the notation, we describe the nature of the problem and the
way we quantify the bias.  In Section \ref{sec:sims} we describe how
we produce simulations which we then use in Section
\ref{sec:bias_synt} to study the size of the bias as a function of the
characteristics of the PSF and of galaxies.  In section
\ref{sec:two_filters} we show that it is possible to construct a
calibration sample to model the bias using observations with two
narrower filters. In Section \ref{sec:HST}, we discuss the performance
one can achieve using observations in the F606W and F814W {\it HST}
filters.  We show in Section \ref{sec:sample}, that the number of
galaxies observed in the F606W and F814W (or F850LP) in the {\it HST}
archive will be large enough to characterise the bias with the
required precision.  We conclude in Section \ref{sec:conclusions}.

\section{Description of the problem}\label{sec:intro}

We start by showing how the spatial variation of the SED across a
galaxy profile affects the shear estimation and introduce the notation
that will be used throughout the paper. To facilitate the readability
of the paper we summarise the main quantities in
Table~\ref{tab:table_moments}.  

Throughout the paper we implicitly assume that measurements are done
on images produced by a photon counting device, such as a
charge-coupled device (CCD). Hence $I(\bx;\lambda)$, the observed
photon surface brightness\footnote{Note that we drop the explicit
  mention of `photon' after the first part of Section~2.} or {\it
  image} at a wavelength $\lambda$, is related to the intensity
$S(\bx;\lambda)$ by $I(\bx;\lambda)=\lambda S(\bx;\lambda)
T(\lambda)$, where $T(\lambda)$ is the normalised transmission. The
image of a galaxy observed with a filter of width
$\Delta \lambda$ is given by

\be
I^{\rm obs}(\bx)=\int_{\Delta \lambda} I^0(\bx;\lambda) \ast P(\bx;\lambda) 
 d\lambda \label{eq:iobs_real}\,,
\ee

\noindent where $P(\bx;\lambda)$ is the wavelength dependent PSF, and
$I^0(\bx,\lambda)$ is the image of the source
before the convolution (denoted by $\ast$) with the PSF. In Fourier space,
Equation.~(\ref{eq:iobs_real}) can be written in a more convenient way as

\be\label{eq:iobs}
I^{\rm obs}(\bk)=\int_{\Delta \lambda} I^0(\bk;\lambda)  P(\bk;\lambda) 
d\lambda \,.
\ee
 
\noindent To measure cosmic shear one needs to estimate the
second-order moments of the PSF-corrected image
$I^{0}(\bx)=\int_{\Delta \lambda} I^0(\bx;\lambda) d\lambda $.  In the
weak lensing regime \citep[see for example][]{BaSc01} the shear can be
estimated from the measurement of second-order moments $Q^0_{ij}$:

\be \tilde \gamma_1 +i \tilde \gamma_2\simeq
\frac{Q^0_{11}-Q^0_{22}+2i
  Q^0_{12}}{Q^0_{11}+Q^0_{22}+2(Q^0_{11}Q^0_{22}-(Q^0_{12})^2)^{1/2}}\,,
\ee

\noindent where we have defined the complex shear $\tilde{\vc
  \gamma}=\tilde \gamma_1+i\tilde \gamma_2$. The second-order moments
of the light distribution are given by

\be\label{eq:moments}
 Q^0_{ij}=\frac{1}{F}\int_{\Delta\lambda} d\lambda\int I^{0}
(\bx;\lambda) x_i x_j d\bx\,, \ee

\noindent where we implicitly assumed that they are evaluated around
the position where the dipole moments vanish. The total observed
photon flux is given by

\be F=\int_{\Delta\lambda} F(\lambda) d\lambda \equiv \int_{\Delta\lambda} d\lambda
\int I^{0} (\bx;\lambda) d\bx\,. \label{eq:fluxdef} \ee

\noindent The observed moments are measured from the PSF-convolved
image given by Equation~(\ref{eq:iobs_real}). In
practice, to reduce the effect of noise in observed images, moments
are evaluated using a weight function $W(|\bx|)$ with a characteristic
size $r_w$. Hence the observed quadrupole moments are given by

\be
Q^{\rm obs}_{ij}=\frac{1}{F_w}\int_{\Delta\lambda} d\lambda \int d \bx I^0(\bx;\lambda)\ast P(\bx;\lambda) x_i x_j  W(|\bx|;r_w)  \,,
\ee

\noindent with $F_w$ indicating the weighted flux (i.e. the weight
function is also introduced in Equation~(\ref{eq:fluxdef})), from which it is
still possible to retrieve the shear (see for example
\citealt{KaSqBr95} or \citealt{Melchior11}).

\begin{table*}
\begin{tabular}{ll}
Symbol & Description \\
\hline
$Q^0_{ij}$ & Second order moments before PSF convolution. \\
   & Without explicit $\lambda$ dependence it refers to the moments integrated over the whole pass-band.  \\
\hline
$I^{0}({\bf x})$ & Photon surface brightness or {\it image} describing the source light distribution before smearing by the PSF.\\
 &  Without explicit $\lambda$ dependence it refers to the counts integrated over the whole pass-band.\\
\hline
$S({\bf x};\lambda)$ & Source intensity as a function of position and wavelength.\\
 & \\
\hline
$Q^{\rm obs}_{ij}$ & Observed second order moments after PSF convolution. \\
 &  Without explicit $\lambda$ dependence it refers to the moments integrated over the whole pass-band. \\
\hline
$I^{\rm obs}({\bf x})$ & Observed photon surface brightness or {\it image}. \\  
&  Without explicit $\lambda$ dependence it refers to the counts integrated over the whole pass-band. \\
\hline
$P_ {ij}({\bf x})$ & Second order moments of the PSF. \\
& Without explicit $\lambda$ dependence it refers to the moments integrated over the whole pass-band. \\
\hline
$R_{\rm PSF}$ & Characteristic size of the PSF estimated as the sum \\
&  of the second order moments: $\sqrt{P_{11}+P_{22}}$\\
\hline
$R_{\rm gal}$ & Characteristic size of the observed galaxy estimated as the sum \\
& of the second order moments: $\sqrt{Q^{\rm obs}_{11}+Q^{\rm obs}_{22}}$\\
\hline
$T(\lambda)$ & Transmission as a function of wavelength $\lambda$.\\
&  \\
\hline
$\tilde {\vc \gamma}$ & Estimate of the complex shear vector. \\
& \\
\hline
$Q^{\rm nograd}_{ij} $ & Second order unconvolved moments of a galaxy  with no colour gradients.\\
& \\  
\hline 
$F(\lambda)$ & Photon flux  at a given wavelength $\lambda$, such that $F=\int F(\lambda)d\lambda=\int \lambda S(\lambda)d\lambda$, where $S$ is the flux density. \\ 
& \\
\hline
${\bf e}^{\rm obs}$ & Observed complex ellipticity. \\
& \\
\hline
$P_\gamma$ & Response of a galaxy ellipticity to a shear ${\bf \gamma}$. $P^{\rm nograd} _\gamma$  is the response of a galaxy that is assumed to have no
colour gradients, \\
&  while $P^{\rm grad} _\gamma$ is the true response.\\
\hline
$I^{0}_{\rm nograd}  ({\bf x})$ & Image constructed assuming
a galaxy has no colour gradients.\\
& This quantity is needed to estimate the bias as described in Section \ref{subsec:measurement}\\
\hline
$I^{\rm obs}_{\rm nograd}  ({\bf x})$ & Observed image obtained after  applying a shear to  $I^{0}_{\rm nograd}  ({\bf x})$ and convolving by the PSF.\\
&   Without explicit $\lambda$ dependence it refers to the moments integrated over the whole pass-band.\\
\hline
$I^{\rm obs}_{\rm grad}  ({\bf x})$ & Observed image obtained after  applying a shear to the true galaxy profile $I^{0} ({\bf x};\lambda)$ and convolution by the PSF.  \\
&   Without explicit $\lambda$ dependence it refers to the moments integrated over the whole pass-band.
\end{tabular}
\caption{\label{tab:table_moments} Summary table of quantities defined in this paper. }
\end{table*}

The alternative to moment-based techniques is to use so-called fitting
techniques
\citep[e.g.][]{im2shape,Kuijken06,Milleretal07,Kitchingetal08,Miller12},
which fit the sheared, PSF-convolved galaxy profiles and thus provide
an estimator of the shear. In this case, the profile itself acts as a
weight and determines the parts of the galaxy profile that are used to
estimate the shear.  The weighting is a very important point to keep in mind
because, as we will show, the amplitude of the colour gradients
induced bias strongly depends on the characteristic scale of the
weight function.
To highlight the importance of the weight function, we first examine
the impact of colour gradients on unweighted moments. We start by
noting that the expression for the quadrupole moments can be written
as

\be\label{eq:q_ij} Q^0_{ij}=\frac{1}{F}\int_{\Delta\lambda} Q^0_{ij}(\lambda)
F(\lambda)d\lambda\,, \ee 

\noindent where  $Q^0_{ij}(\lambda)$ is given by Equation (\ref{eq:moments}) with an infinitesimally narrow  filter centered on $\lambda$. For the following it is convenient to define $\lambda_{\rm ref}$, 
a reference wavelength such that $F\equiv F(\lambda_{\rm
  ref})\Delta\lambda$. Using a second order Taylor expansion for both
$F(\lambda)$ and $Q^0_{ij}(\lambda)$ around $\lambda_{\rm ref}$, and
keeping only the even powers (the odd powers will vanish after
integration) one finds that

\ba Q^0_{ij}=\frac{1}{F}\int_{\Delta\lambda} F(\lambda_{\rm ref})
Q^0_{ij}(\lambda_{\rm ref})+\frac{1}{2} f_2 Q^0_{ij}(\lambda_{\rm ref})
(\lambda-\lambda_{\rm ref})^2 d\lambda\label{eq:colour}\\ + \frac{1}{F}
\int_{\Delta\lambda}
\Big(f_1 q_{ij,1} +\frac{1}{2} q_{ij,2} \Big) (\lambda-\lambda_0)^2
d\lambda + {\mathcal O}(\Delta \lambda^4) \nonumber\,, \ea 

\noindent where we have defined 

\ba f_k=\frac{\partial^k F(\lambda)}{\partial
  \lambda^k}\Big|_{\lambda_{\rm ref}}\nonumber\,, {\rm ~and}~~~q_{ij,k}=\frac{\partial^k
  Q^0_{ij}(\lambda)}{\partial \lambda^k}\Big|_{\lambda_{\rm ref}}\,.\nonumber
\ea

\noindent We have split the right-hand side of Equation
(\ref{eq:colour}) into two terms on purpose. The first term
corresponds to the second-order moments if the galaxy had no colour
gradients. The second term represents the lowest order correction due
to colour gradients. By evaluating the integral we obtain:

\ba\label{eq:colors1} Q^0_{ij}=Q^{\rm nograd}_{ij}+\frac{(\Delta\lambda)^2}{12}
\Big({\frac{q_{ij,1} f_1}{F(\lambda_{\rm ref})}+\frac{1}{2}q_{ij,2}}\Big)
+ {\mathcal O}(\Delta \lambda^4)\,, \ea

 \noindent which shows that the change in the quadrupole moments 
is proportional to $(\Delta \lambda)^2$. Note that

\be\label{eq:colors2}
\frac{f_1}{F(\lambda_{\rm ref})}\equiv\frac{1}{F(\lambda_{\rm ref})}
\frac {\partial F(\lambda)}{ \partial \lambda}\Big|_{\lambda_{\rm ref}}= \frac{\partial \ln F(\lambda)}{\partial \lambda }\Big|_{\lambda_{\rm ref}}\,,
\ee

\noindent which means that the change of the moments depends on the
variation of the SED of the galaxy across the filter, i.e. on its colour.

Interestingly, it is still possible to determine $Q^0_{ij}$
from the observed {\it unweighted} quadrupole moments. This can
be seen by writing down the convolution explicitly and changing
the order of integration. Doing so we obtain \citep[also see][]{Valdes83}:

{\setlength\arraycolsep{0.1em} \ba\label{eq:unweighted} Q^{\rm
    obs}_{ij}&=&\frac{1}{F}\int I^{0}(\bx;\lambda)\ast P(\bx;\lambda)
  x_i x_j d\lambda\nonumber\\ &=&Q^{0}_{ij}+\frac{1}{F} \int
  F(\lambda) P_{ij}(\lambda) d\lambda\,.  \ea }

\noindent The observed quadrupole moments can thus be written as the
sum of the true moments and $P_{ij}(\lambda)$, the quadrupole moments
of the PSF integrated over the pass-band. More generally, even moments
of order $N$ of $I^{0}(\bx)$ are expressions involving the moments of
$I^{0}(\bx;\lambda)$ up to $N-2$. In the case of the second-order
moments, only the knowledge of the SED is needed to retrieve
$Q^0_{ij}$. Hence colour gradients do not bias the shear estimate
based on unweighted moments, provided the SED of the galaxy and the
PSF moments as function of the wavelength, $P_{ij}(\lambda)$, are
known.

In contrast, it is not possible to correct for the PSF without
knowledge of the higher order moments when weighted quadrupole moments
are used. It is therefore not possible to recover $I^{0}(\bx)$, 
except in two special cases. The first case is when the PSF is achromatic: 

\be I^{0}(\bk)
=\frac{1}{P(\bk)}\int_{\Delta\lambda} d\lambda{I^{\rm obs}(\bk;\lambda)}
=\frac{I^{\rm obs}(\bk)}{P(\bk)}\,.  \ee

\noindent The second case is when the galaxy has no colour gradients:

\be 
I^{0}(\bk;\lambda_{\rm ref})=\frac {F(\lambda_{\rm ref}) I^{\rm  obs}(\bk)}
{\int_{\Delta\lambda} F(\lambda)P(\bk;\lambda)d\lambda}\label{eq:PSF}\,,  \ee

\noindent and one can rewrite $I({\bf x}; \lambda) =
I^{0}({\bk};\lambda_{\rm ref})F(\lambda)/F(\lambda_{\rm ref})$ 
for any given choice of $\lambda_{\rm ref}$.  In this last
case one is able to derive the PSF corrected image by knowing both
the flux of the galaxy and the PSF profile as a function of
wavelength. This is the case studied by \citet{Cyprianoetal10}.

\subsection{Measurement}\label{subsec:measurement}

The need to use a weight function when measuring the shapes of the
faint galaxies  leads to a bias
in shear estimates due to colour gradients. This result applies to all
moment-based methods, such as KSB \citep{KaSqBr95} or DEIMOS
\citep{Melchior11}. However, the fact that one cannot recover the
unconvolved image $I^{0}(\bx)$ from broad-band observations suggests
that `fitting methods' will also be prone to provide biased estimators
of the shear unless they are able to account for the existence of
colour gradients. For instance, one could attempt to model galaxies
with a bulge and disk component, each with their own SED. We proceed
to quantify this bias for moment-based methods, but note that our
approach can be extended to evaluate the bias for fitting techniques
as well.

The first step is to define the bias that is induced by a spatially
varying SED. The PSF correction described by Equation (\ref{eq:PSF})
is perfect for a galaxy that has no colour gradients. If Equation
(\ref{eq:PSF}) is used to obtain the unconvolved image of a galaxy,
one effectively approximates the observed galaxy with a galaxy that
has the same SED, but no colour gradients.  This galaxy actually has a
different profile and thus its response to the shear and to the PSF is
different. For this reason, correcting the galaxy using Equation
(\ref{eq:PSF}) leads to a biased estimate of the shear. To quantify
the resulting bias we define the response $P_{\gamma}$ as the link
between the observed ellipticity ${\bf e}^{\rm obs}$ and the shear
${\vc \gamma}$

\be\label{eq:estimator} {\bf e}^{\rm obs}=P_\gamma {\vc
  \gamma}\,, \ee

\noindent where we defined the observed complex ellipticity using
weighted quadrupole moments:

\be
 e^{\rm obs}_1 +i e^{\rm obs}_2= \frac{Q^{\rm obs}_{11}-Q^{\rm obs}_{22}+2i Q^{\rm obs}_{12}}{Q^{\rm obs}_{11}+Q^{\rm obs}_{22}+2(Q^{\rm obs}_{11}Q^{\rm obs}_{22}-(Q^{\rm obs}_{12})^2)^{1/2}}\,.\label{eq:defellipticity}
\ee 

\noindent In the absence of observational biases, Equation
(\ref{eq:estimator}) can be used to obtain an unbiased estimate of the
shear. In practice the shear is obtained by averaging over many
galaxies, since they have an intrinsic ellipticity that is much larger
than a typical shear. In the case of a galaxy with colour gradients
one is unable to estimate $P^{\rm grad}_{\gamma}$, the correction to
apply to obtain an unbiased shear estimate. One will instead correct
the PSF using Equation (\ref{eq:PSF}) which replaces the response
$P_{\gamma}^{\rm grad}$ with $P_{\gamma}^{\rm nograd}$. This
approximation leads to a multiplicative bias which we define as:

\be m\equiv\frac{\tilde \gamma_i} {\gamma_i}-1=\frac{P_{\gamma}^{\rm
    grad}}{P_{\gamma}^{\rm nograd}}-1\,.  \ee

\noindent where $\gamma_i$ refers either to the first or second
component of the shear pseudo-vector; we make the assumption that the
bias $m$ is the same for both components. To estimate the difference
in response we create a pair of galaxies which appear identical when
observed through a broad pass-band. One of them has a colour gradient
and the other does not. We apply the same shear to both galaxies,
convolve them with their respective PSFs and derive $P_{\gamma}^{\rm
  nograd}$ and $P_{\gamma}^{\rm grad}$ by comparing the observed
ellipticities to the applied shear.
 
\begin{figure}
\begin{tabular}{|@{}l@{}|@{}l@{}|}
\psfig{figure=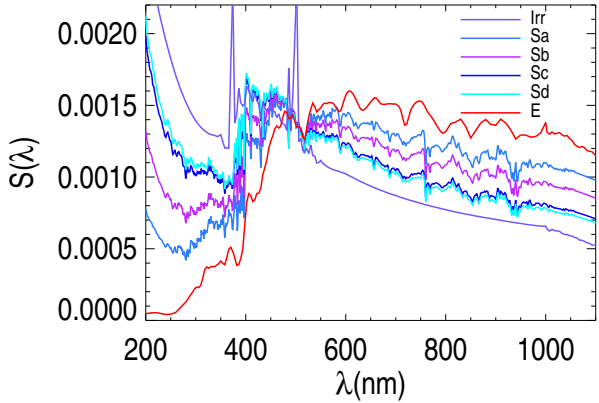,width=0.5\textwidth}
\end{tabular}
\caption{ \label{fig:SED} Spectral energy distributions used to create
  the disk and bulge components of our synthetic galaxies. All SEDs
  are normalised such that the integrated $S(\lambda)$ between
  $200$ and $1100$ nm is one.  In our synthetic galaxies the SED of
  the bulge (red solid line) is fixed, while the SED of the disk can
  be either Sa, Sb, Sc, Sd simulating a bluer and bluer galaxy, or Irr
  which represents a starburst population.  All reference galaxies
  (see Table \ref{tab:1}) use an Irr SED.}
\end{figure} 
  
\section{Simulations}\label{sec:sims}

\begin{figure*}
\begin{tabular}{|@{}l@{}|@{}l@{}|}
\psfig{figure=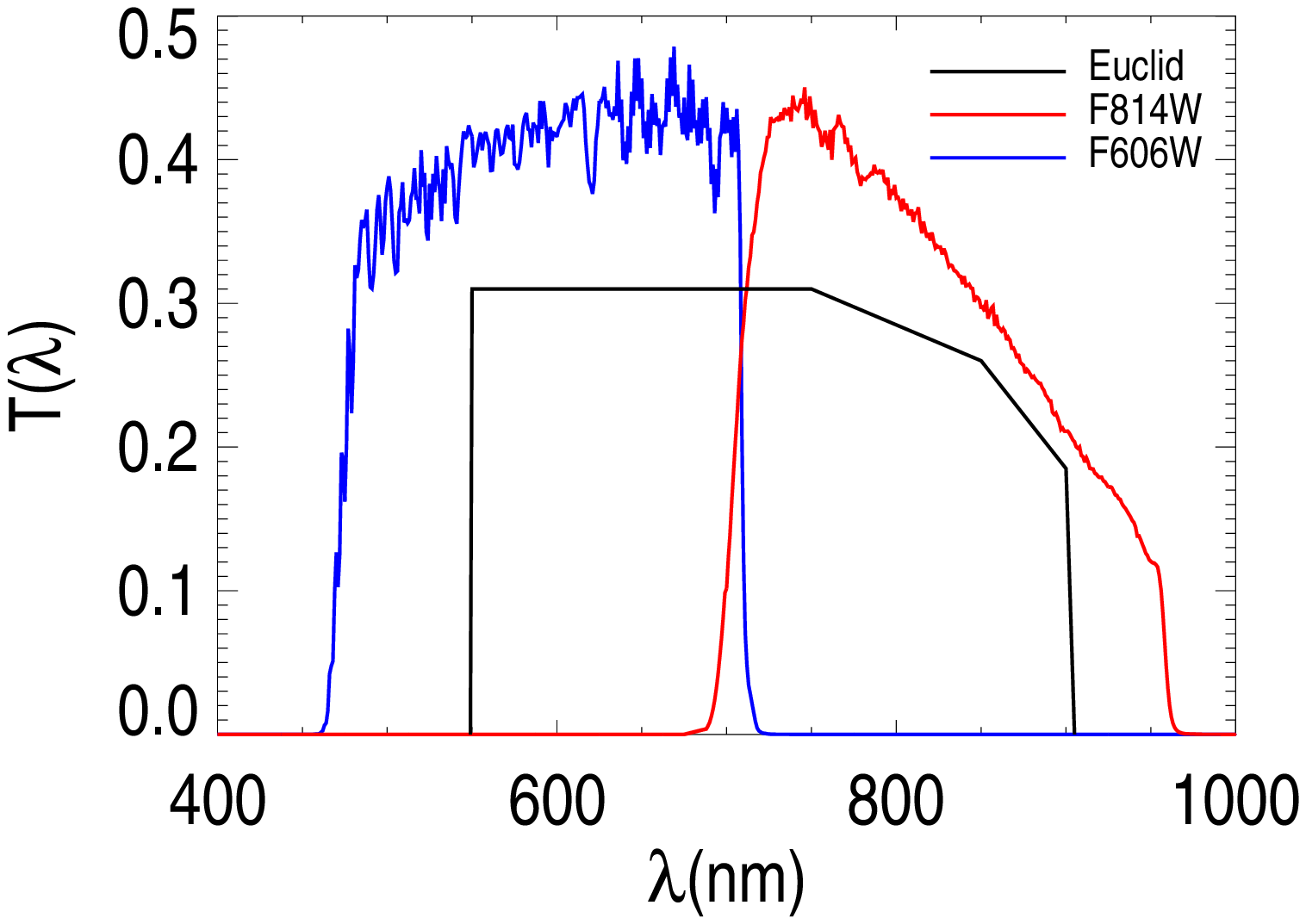,width=0.5\textwidth}&\psfig{figure=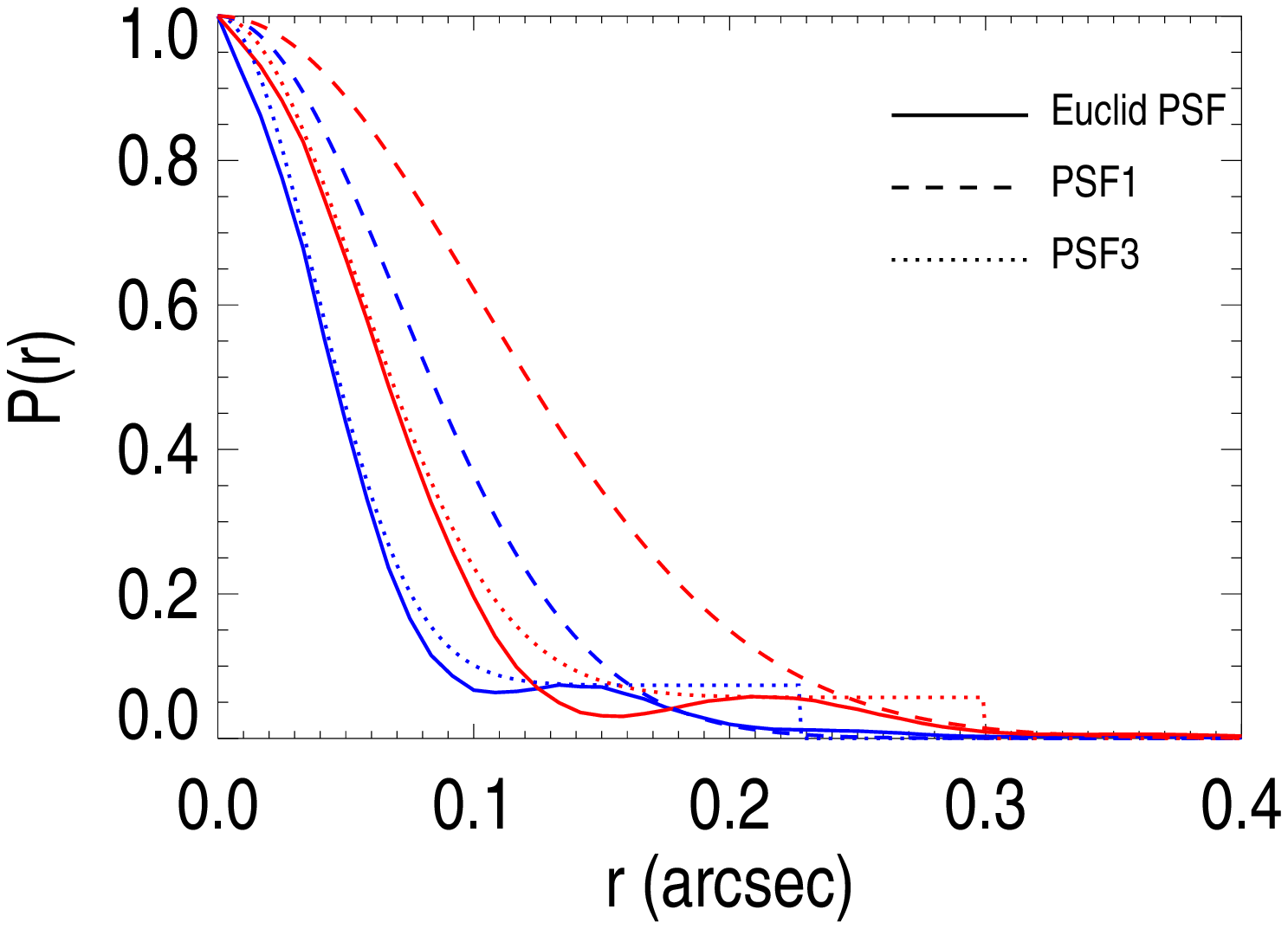,width=0.5\textwidth}
\end{tabular}
\caption{\label{fig:filters} Left panel: The adopted {\it Euclid}
  transmission curve $T(\lambda)$ (black solid line) and those for the
  {\it HST} filters F606W (blue solid line) and F814W (red solid line)
  used in Section \ref{sec:two_filters} and \ref{sec:HST} to model the
  bias. Note that the normalisation of the {\it Euclid} transmission
  curve is arbitrary, whereas for the F606W and F814W filters we show
  the measured throughput $T(\lambda)$.  Right panel: comparison of
  the {\it Euclid} PSFs at $550~ {\rm nm}$ (solid blue line) and $800~
  {\rm nm}$ (solid red line) and our reference model PSF (PSF1; dashed
  blue and red lines). The parameters for our model PSFs are listed in
  Table \ref{tab:PSFs}. PSF1 is broader and has a stronger wavelength
  dependence than the actual {\it Euclid} PSF.  The dotted lines show
  the profiles for PSF3, which resembles the {\it Euclid} PSF more
  closely (see also Section~\ref{subsec:FWHM}).}
\end{figure*} 

The main aim of this paper is to develop a method that can be used to
estimate the bias induced by the presence of colour gradients.  In a
future paper (Huang et al. in prep.) we will apply our approach to
determine the bias using real data. Here we examine instead whether it
is possible in principle to measure the colour gradient bias with
sufficient precision for {\it Euclid}. We do so by making a number of
conservative assumptions, while using synthetic galaxies and an
analytic description of the PSF.

We assume that galaxies can be described as the sum of a bulge and a
disk component, each characterised by S\'ersic profiles of index $n$.
The S\'ersic profile of index $n$ is given by:

\be
S(\bx)=S_c e^{-\kappa[(\bx-\bx_0)^T {\bf C} (\bx-\bx_0)]^{\frac{1}{2n}}  }
\ee

\noindent where ${\bf x}^T$ is the transpose of ${\bf x}$,
$S_c$ is the value of the intensity in the center $\bx_0$,
$\kappa=1.9992n-0.3271$ (see for example \citealt{Capaccioli89}),
and ${\bf C}$ is a matrix defined by:

{\setlength\arraycolsep{0.1em}
\ba
C_{11}&=&\Big(\frac{\cos^2(\phi)}{a^2} + \frac{\sin^2(\phi)}{b^2} \Big)\,,\\
C_{12}&=&\frac{1}{2}\Big(\frac{1}{a^2} + \frac{1}{b^2}\Big)\sin(2\phi)\,,\\
C_{22}&=&\Big(\frac{\sin^2(\phi)}{a^2} + \frac{\cos^2(\phi)}{b^2} \Big) \,,
\ea
}

\noindent with $\phi$ the angle between the semi-major axis $a$ and
the x-axis and $b$ the semi-minor axis. In the case of an axisymmetric
profile $r_h=a=b$ is the half-light radius. The ellipticity of the
profile is defined as $e=(a-b)/(a+b)$.

We acknowledge that this double S\'ersic model may not describe the
full variety of observed galaxies. As our aim is not to determine the
bias, but rather develop an approach to measure the bias,  this model is sufficient for our purposes as it
should capture most of the problem of colour gradients. Whenever statements which depend on the value of the bias have to be made, we will make
sure that our assumptions are conservative and the bias is not
underestimated. 

To describe the SED of the bulge and the disk we use the galaxy
templates from \citet*{Coetal80} shown in Figure \ref{fig:SED}. The
bulge is always modelled with the SED of an old stellar population
typical of elliptical galaxies, while the disk is modelled either with
an extremely blue SED typical of an irregular galaxy, or intermediate
stellar populations such as Sa, Sb, Sc and Sd. For each wavelength we
construct the galaxy surface brightness profile $S^0(\bx;\lambda)$ by
adding up the profiles of the bulge and the disk normalised by fixing
the ratio $S_{\rm bulge}/S_{\rm disk}$ at $\lambda=550~ {\rm nm}$. The
profile is sampled in a wavelength range between $200$ and $1100$ nm
with a step of $1$ nm. The pixel size of each $S^0(\bx;\lambda)$ image
is $0.05~{\rm arcsec}$.

In order to generate the observed synthetic galaxy image we need to
decide on the shape of the transmission curve $T(\lambda)$ and the
profile of the PSF at any given wavelength. The black solid line in
the left panel of Figure \ref{fig:filters} shows the curve we use to
model the expected transmission for {\it Euclid}. Note that we define here as {\it Euclid} PSF, the PSF used in the Design  Study \citep{redbook}. This PSF is not the final one but we do not expect major changes.
 We use as reference
a PSF with a Gaussian profile and wavelength dependent width:

\be\label{eq:PSF_dispersion} %% write it down in arcsecs....
\sigma(\lambda)=w_{0,800} \Big(\frac{\lambda}{800{\rm nm}}\Big)^{\alpha},
\ee

\noindent where $\alpha$ is the slope and $w_{0,800}$ is the width of the PSF at
$800 ~{\rm nm}$.  As discussed by \cite{PaHeetal09} and
\cite{Masseyetal12}, the ability to recover the shear of a galaxy of
given characteristic size $R_{\rm gal}^2=Q_{11}^{\rm obs}+Q^{\rm
  obs}_{22}$ is $\propto R_{\rm PSF}^2/R_{\rm gal}^2$ where $R_{\rm
  PSF}^2=P_{11}+P_{22}$ is the characteristic size of the PSF. This
reflects the fact that the PSF blurs the galaxy and this effect can be
only partially corrected.  

We construct our reference PSF by choosing $\alpha=1$ and
$w_{0,800}=0.102~{\rm arcsec}$. This yields a PSF with the same
characteristic size $R_{\rm PSF}^2$ as the {\it Euclid} PSF for 
$\lambda=800 ~{\rm nm}$ but with a stronger wavelength dependence (see
right panel of Figure \ref{fig:filters}). For comparison purposes we
also define an even broader PSF, indicated as PSF2 in
Table~\ref{tab:PSFs}. The {\it Euclid} PSF is not well approximated by
a single Gaussian, but instead can be considered as being composed of
two parts: the core with a diffraction limited characteristic size
$\propto \lambda$ and the wings which also contribute to $P_{ij}$ but
have a much weaker wavelength dependence. For this reason the overall
{\it Euclid} PSF size scales as $R_{\rm PSF}^2\propto
\lambda^{0.55}$. Its FWHM does scale proportionally to $\lambda$ as it
depends on the core.

Our reference PSF is much broader than the {\it Euclid} PSF and this
leads to bias estimates considerably larger than the ones expected for
the actual {\it Euclid} PSF.  To quantify how conservative the results
are when using the reference PSF, we compare to PSF3 which is the sum of
a Gaussian and a top-hat. The Gaussian, with a width
$w_{0,800}=0.054\,{\rm arcsec}$, is chosen to fit the core of the {\it
  Euclid} PSF, while the top-hat approximates the wings. The cut-off
size of the top-hat is approximately at the position of the second
minimum of the PSF profile, $r_{\rm cut-off} \propto \lambda^{0.74}$
and the normalisation is such that the top-hat contains ~$20\%$ of the
total flux. As shown in the right panel Figure \ref{fig:filters}, PSF3
approximates the main properties of the {\it Euclid} PSF fairly well.

\begin{table}
\begin{tabular}{ll}
PSF & Description \\
\hline
{\bf Reference} (PSF1) &  Gaussian PSF described by Eqn. (\ref{eq:PSF_dispersion})\\
& with $w_{0,800}=0.102\, {\rm arcsec}$  and $\alpha=1$.\\  
\hline
Wide Gaussian (PSF2)  & Gaussian PSF described by Eqn. (\ref{eq:PSF_dispersion}), \\
&  with $w_{0,800}=0.15\, {\rm arcsec}$  and $\alpha=1$.   \\       
\hline
Gaussian + top-hat (PSF3) & Gaussian core described by Eqn. (\ref{eq:PSF_dispersion})\\     
&  with $w_{0,800}=0.054~ {\rm arcsec}$  and $\alpha=1$ \\
&  and top-hat with 20\% of the total flux   \\
& and a cut-off size  $\propto \lambda^{0.74}$.    \end{tabular}
\caption{\label{tab:PSFs} Summary table of the PSFs considered in
  Section \ref{subsec:FWHM}. The reference PSF is used throughout the
  paper, as its large width exacerbates the bias. PSF3 resembles the
  actual {\it Euclid} PSF more closely.}
\end{table}

Using the model for the galaxy, the PSF and the transmission, we
compute the observed image $I^{\rm obs}(\bx)$. We proceed following
the procedure outlined in Section~\ref{sec:intro} and create pairs of
galaxies that appear identical in the broad-band image, but where one
has a colour gradient and the other does not.  To create the latter,
we deconvolve $I^{\rm obs}(\bx)$ using the colour-weighted PSF
correction in Equation (\ref{eq:PSF}) and call this $I^{0}_{\rm
  nograd}(\bx;\lambda)$. We apply the same shear to
$I^{0}(\bx;\lambda)$ and $I^{0}_{\rm nograd}(\bx;\lambda)$. We
convolve the profiles by the PSF $P(\bx;\lambda)$ and create the final
observed images by summing $I^{\rm obs}_{\rm grad}(\bx;\lambda)$ and $I^{\rm obs}_{\rm nograd}(\bx;\lambda)$
over the full pass-band. Note that we use a pixel size of $0.05\, {\rm
  arcsec}$, because we do not want undersampling to affect the
estimate of the bias. To construct the $I^{\rm obs}_{\rm
  nograd}$ and $I^{\rm obs}_{\rm grad}$ images one would want the best
resolution available to avoid sampling bias. The sampling we use
corresponds to the {\it HST} ACS/WFC resolution, which is sufficient
to avoid undersampling of the {\it Euclid} PSF. We then measure the
ellipticities from these images using weighted quadrupole moments and
determine $P^{\gamma}$ for each galaxy. To reduce noise in our
estimate of the multiplicative bias we use the `ring-test' method
\citep{NaBe07} creating 8 copies of the same galaxy but with different
orientation.

\begin{figure}
\begin{tabular}{|@{}l@{}|@{}l@{}|} 
\psfig{figure=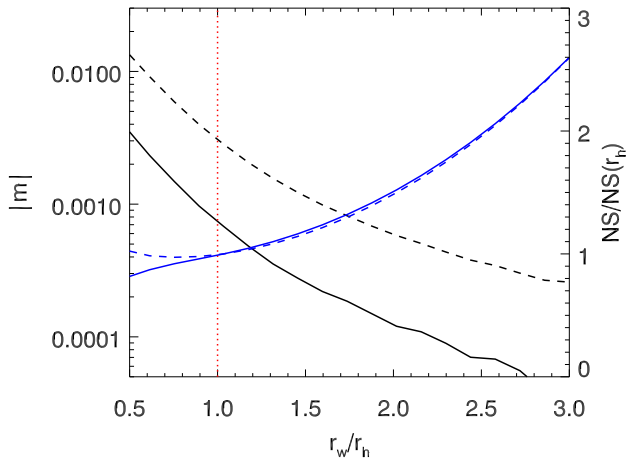,width=0.5\textwidth}
\end{tabular}
\caption{\label{fig:effect_weight} Amplitude of the absolute value of
  the bias as a function of $r_w$, the width of the weight function,
  in units of $r_h$, the half-light radius, for the reference galaxies
  B (solid black line) and S (dashed black line) defined in Table
  \ref{tab:1} for $z=0$. The dotted red line indicates where the weight function
  size is equal to $r_h$. The blue solid (dashed) line indicates the
  noise-to-signal ratio for the B (S) galaxy as a function of the filter
  size normalised to the value obtained when $r_w=r_h$. Note that the
  noise-to-signal ratio shown here only includes the contribution from the
  sky background (see text).}
\end{figure}

\begin{table*}
\begin{tabular}{ccccccc}
 Name   &  SED  & a (arcsec) & Flux ratio (550 nm) & S\'ersic index & FWHM (z=0)(arcsec) & FWHM(z=0.9) (arcsec) \\ 
\hline
B  & E/Irr & 0.17/1.2  &   $25\%/75\%$ & 1.5/1.0 & 0.32 & 0.40 \\
S  & E/Irr  & 0.09/0.6  &   $25\%/75\%$ & 1.5/1.0 & 0.27 &  0.37 \\
B4 & E/Irr   & 0.39/0.35 &   $33\%/67\%$ & 4.0/1.0 & 0.22 & 0.30 \\
S4 & E/Irr & 0.12/0.20 &   $33\%/67\%$ & 4.0/1.0 & 0.22 & 0.27 \\
\end{tabular}

\caption{\label{tab:1}Characteristics of our reference galaxies. When
  two values are quoted in a column, the first value refers to the
  bulge the second to the disk. The galaxies are circular ($a=b$) with
  the size of the semi-major axis $a$ indicated in the second column.
  The observed FWHM at $z=0$ and $z=0.9$ are computed using the
  reference Gaussian PSF.  Note that for these galaxies we keep the
  S\'ersic parameters of disk and bulge fixed as a function of
  redshift.  As a result, the FWHM increases at high redshifts because
  the disk becomes brighter than the bulge. A second pair of galaxies
  (B4 and S4) is chosen such that the bulge-to-disk properties are
  similar to the typical galaxy from \citet{Voigtetal12} and
  \citet{Simardetal02}.}
\end{table*}

\section{Evaluation of the bias with synthetic galaxies}\label{sec:bias_synt}

The intrinsic properties of the source galaxies vary with redshift,
and the colour gradient bias will therefore differ between tomographic
redshift bins. As shown below \citep[but also see][]{Masseyetal12} the
bias depends on the intrinsic galaxy size and for this reason we
define a number of model galaxies in Table~\ref{tab:1} that span a
range in size. The properties of these galaxies are chosen purposely
to have large colour gradients: they are a superposition of a rather
bright red bulge with small characteristic size and an extended disk
with an SED of an irregular galaxy.

We note that the reference PSF used to compute the observed FWHM
values listed in Table~\ref{tab:1} is much broader than the {\it
  Euclid} PSF (see Fig.~\ref{fig:filters}). As discussed in
\cite{redbook} the {\it Euclid} source galaxy sample is selected to
have an observed FWHM $>1.25 \times {\rm FWHM_{\rm PSF}}$ (where ${\rm
  FWHM_{\rm PSF}}\sim 0.15''$ is the FWHM of the PSF at $\lambda=800\,
{\rm nm}$). Comparison to observed sizes in deep {\it HST} data
indicate that galaxy S is actually representative for the
smallest galaxies that will be used in the {\it Euclid} weak lensing
analysis \citep{Masseyetal12}.

\subsection{Dependence on the weight function}

As mentioned in Section \ref{sec:intro}, the bias is a function of the
weight function, which we take to be a Gaussian with a dispersion
$r_w$. The optimal choice, in terms of maximising the signal-to-noise
ratio, is to match the weight function to the size of the source
galaxy. Figure~\ref{fig:effect_weight} shows the amplitude of the bias
as a function of $r_w$ for the B and S model galaxies defined in
Table~\ref{tab:1}. For the reference galaxy B we obtain a bias of
approximately $-8 \times 10^{-4}$ if we take $r_w$ to be equal to the
half-light radius $r_h$, which was measured on the original observed
image $I^{\rm obs}_{\rm grad}(\bx)$ (i.e., without applying
shear). Compared to the B galaxy, which has the same bulge-to-disk
flux ratio, the colour gradient-induced bias is larger for the S
galaxy because its bulge and disk sizes are both a factor of two
smaller.  In this case the bias is $-3 \times 10^{-3}$ when $r_w=r_h$.

\begin{figure}
\begin{tabular}{|@{}l@{}|@{}l@{}|} 
\psfig{figure=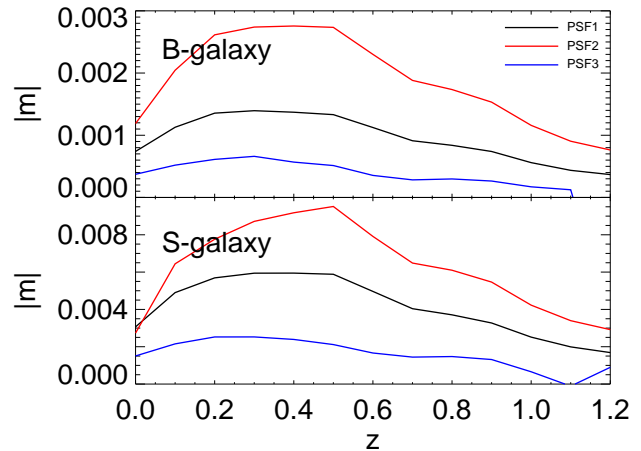,width=0.5\textwidth}
\end{tabular}
\caption{\label{fig:effect_size} The top panel shows the value of the bias for the
  B galaxy as a function of redshift for the PSFs in Table
  \ref{tab:PSFs}. The black solid line shows the result for the
  reference PSF (PSF1); the red solid line shows the bias for a
  Gaussian with a larger width (PSF2), while the blue line indicates
  the bias when the PSF is approximated by a Gaussian with width
  $w_{0,800}= 0.05~{\rm arcsec}$ plus a top-hat (PSF3) whose profile
  is shown in right panel of Figure \ref{fig:filters}. The bottom
  panel shows the same results but for the S galaxy. }
\end{figure} 

The bias increases when $r_w$ decreases, whereas the bias vanishes
when $r_w$ goes to infinity, as expected from Section
\ref{sec:intro}. This result can be understood by noting that the PSF
correction described by Equation (\ref{eq:PSF}) is weighted by the
colour of the galaxy. For galaxy B, the disk contains $75\%$ of the
flux whereas the bulge contains only $25\%$ of the total flux. When
$r_w$ is small, the moments are measured essentially from the profile
of the bulge, and corrected using an effective colour of the PSF which
is always wrong. For the disk the colour-weighted PSF is closer to the
correct one and therefore the bias is reduced when $r_w$ increases.

\begin{figure*}
\psfig{figure=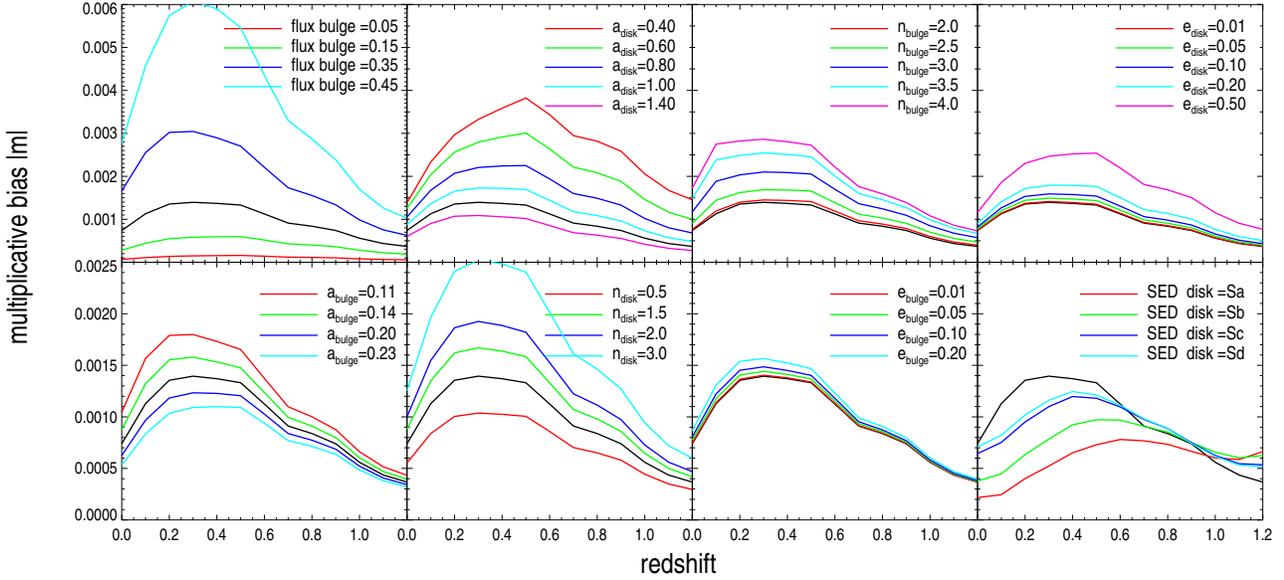,width=1.0\textwidth}
\caption{\label{fig:parameters} Amplitude of the absolute value of the
  multiplicative bias for the B galaxy described in Table \ref{tab:1}
  as a function of redshift, when varying the bulge and disk
  parameters and using PSF1 (see Table~\ref{tab:PSFs}).  Note that the
  S\'ersic parameters of the disk and bulge are fixed as a function of
  redshift. For comparison, in each panel the black solid
  line indicates the bias for the reference galaxy B.  Top panels: from left to
  right we vary the percentage of the light in the bulge, the
  semi-major axis value of the disk (in arcsec), $a_{\rm disk}$, the
  S\'ersic index of the bulge, the ellipticity of the disk. In the
  bottom panels we vary: the characteristic size of the bulge $a_{\rm
    bulge}$ (in arcsec), the S\'ersic index of the disk, the
  ellipticity of the bulge, the SED of the disk.}
\end{figure*}  

Throughout the rest of the paper we will choose the value for $r_w$ to be equal
to the observed galaxy half-light radius $r_h$ because this is the
optimal choice in terms of signal-to-noise ratio, and therefore
routinely used in practice. This choice, however, may no longer be
optimal when colour gradient-induced biases are considered. To show
this, we compare the value of the bias with the noise-to-signal ratio
of the observed ellipticity as a function $r_w$. We
assume that the pixel noise is dominated by the sky background, which
is the case for the majority of galaxies in a typical weak lensing
survey. 

The blue lines in Figure~\ref{fig:effect_weight} indicate the
noise-to-signal ratio in the measurement of the ellipticity as a
function of $r_w$. Since we are only interested in the
relative change, the noise-to-signal ratio is normalised to its value
when $r_w=r_h$. Because we ignored the contribution to the noise that
depends on the galaxy light profile (see Appendix A of
\citealt{HoFrKu00}) the ratio is approximately constant for small
$r_w$. Including this term would result in an upturn in the
noise-to-signal ratio for small $r_w$, such that taking $r_w\sim r_h$
minimises the noise-to-signal ratio. Figure~\ref{fig:effect_weight}
shows that the colour-gradient bias decreases rapidly when $r_w$ is
increased. The noise-to-signal ratio increases but does so relatively slowly
(note the logarithmic scale for the bias and the linear scale for the
noise-to-signal ratio). This suggests that adopting $r_w>r_h$ might
provide a good compromise between reducing the amplitude of the
bias and increasing the noise-to-signal when one needs to account for
colour gradients.

\subsection{Dependence on the PSF and galaxy characteristics}
\label{subsec:FWHM}

In this section we study how the bias depends on the PSF
characteristics. The bias as a function of redshift for the two
reference galaxies B and S is shown in Figure~\ref{fig:effect_size}.
The black solid line indicates the result obtained using the reference
PSF (PSF1 in Table \ref{tab:PSFs}). The red solid line shows the
result for PSF2, which has a larger width resulting in an increase in
the bias. More realistic estimates for the actual bias for {\it
  Euclid} are obtained using PSF3, indicated by the blue lines in
Figure~\ref{fig:effect_size}; the bias is a factor $\sim 2$ smaller
compared to the results for our reference PSF.

%Since the bias depends on the square of the the ratio of the PSF to
%galaxy size \citep[see e.g.][]{Massey12}, future large ground-based
%surveys will also be sensitive to colour gradients, even though they
%typically employ narrower filters and the wavelength dependence of the
%PSF is weaker. 

%One should note that since the bias depends on squared of the the
%ratio of the PSF to the galaxy size, future large ground-based surveys
%might also be sensitive to it. In fact, atmospheric refraction {\bf
%  changes the direction of blue light} more than red light; as a
%result, even ground-based PSFs are wavelength dependent although this
%dependence is weaker than for space-based diffraction limited
%observations.  On the other hand, ground-based PSFs are much larger
%than the space-based ones. For instance, the PSF of the Large Synoptic
%Survey Telescope (LSST) has a size a factor of {\bf five} larger than
%the {\it HST} PSF and because of that, one might require a mitigation
%scheme for the colour gradient induced bias.

The results presented in Figure~\ref{fig:effect_size} are obtained by
changing the redshift, but keeping the input S\'ersic parameters of
the galaxies fixed. This leads to a change in the observed FWHM because
the disk becomes brighter than the bulge as the redshift increases.
Since the simulated galaxies have rather extended disks their observed
sizes also increases (see the comparison of the FWHM at $z=0$ and
$z=0.9$ listed in Table~\ref{tab:1}). Hence, the value of the bias
changes with redshift in part because the ratio of the galaxy size to
the PSF size changes.

Many parameters contribute to the actual spatial colour variations
and  it is therefore useful to examine how the bias changes as a
function of the bulge and disk characteristics. To do so, we take
galaxy B and vary one parameter at a time, keeping all others
constant (using PSF1 and adopting $r_w=r_h$).  The results are
presented in Figure \ref{fig:parameters}. We find that the bias
depends most strongly on the flux of the bulge, but also depends on
the size of both bulge and disk. It does not depend much on their
ellipticity .

\begin{figure}
\begin{tabular}{|@{}l@{}|@{}l@{}|}
\psfig{figure=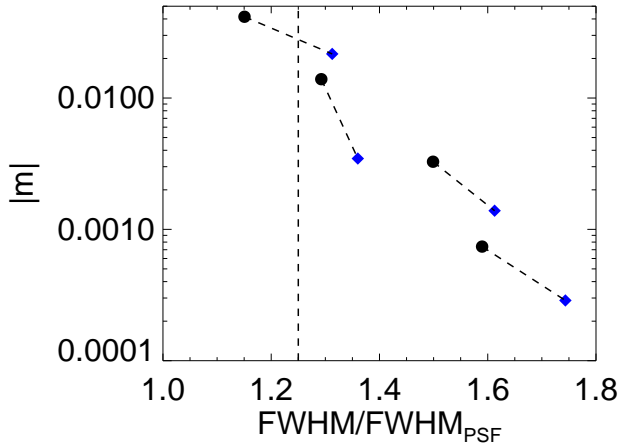,width=0.5\textwidth}
\end{tabular}
\caption{\label{fig:size} Absolute value of the multiplicative bias as
  a function of the ratio of the galaxy FWHM to the FWHM of the PSF
  for the model galaxies S4, B4, S and B (from left to right). The
  results are for $z=0.9$, which corresponds to the median redshift of
  {\it Euclid}. The black points indicate results for the reference
  Gaussian PSF (PSF1 in Table \ref{tab:PSFs}) for which ${\rm
    PSF_{FWHM}}=0.241\, {\rm arcsec}$. Blue values show the results
  for PSF3 (${\rm PSF_{FWHM}}=0.130\, {\rm arcsec}$), which is a
  better approximation of the {\it Euclid} PSF. Dotted lines link the
  different results obtained for the same source galaxy when changing
  the PSF description. The dashed line indicates the limit below which
  galaxies are considered too small to be used in the {\it Euclid} weak
  lensing analysis.}
\end{figure}

In general the bias values range between values of a few times
$10^{-4}$ to a few times $10^{-3}$. Note that in all panels the
rest-frame colour of the galaxies is the same, except when we vary the
fraction of the flux in the bulge (top left panel) or when we change
the SED (bottom right panel). These results confirm what we concluded
based on Equations (\ref{eq:colors1}, \ref{eq:colors2}), i.e. the bias
is foremost a function of the colour of the galaxy. Whereas the
absolute value of the bias increases when the size of the source
galaxy decreases, the changes as a function of the parameters are
similar for all galaxies in Table \ref{tab:1}. The reference galaxy
and its smaller version both have a relatively large disk, which leads
to a large FWHM at $z=0.9$ (see Table~\ref{tab:1}). Therefore we also
consider two galaxies with a smaller FWHM: B4 and S4. The ratio ${\rm
  FWHM/FWHM_{PSF}}$ at $z=0.9$ is 1.3 and 1.1 for B4 and S4,
respectively. As one can see from Figure \ref{fig:size}, the bias is a
very strong function of this ratio: it is smaller than $10^{-3}$ for a
well-resolved galaxy and can become a few percent for a galaxy which
is about the size of the PSF.

Figure~\ref{fig:size} also shows that for all galaxies in Table
\ref{tab:1} the bias is larger for the reference Gaussian PSF (black
points) than for the more realistic PSF3 (blue diamonds). However, the
behaviour of the bias as a function of ${\rm FWHM/FWHM_{PSF}}$ is very
similar for the two PSF models. In particular, the bias is still a few
percent when ${\rm FWHM} \simeq 1.25\,{\rm FWHM}_{\rm PSF} $. Note,
however, that in the case of a PSF as small as that of {\it Euclid} we
do not expect galaxies with observed sizes similar to the PSF
\citep[e.g.][]{Masseyetal12}. For this reason, a characteristic size of about
1.4 ${\rm FWHM}_{\rm PSF}$ can be considered representative for the
{\it Euclid} galaxy sample. For such galaxies we expect a bias of a
few times $10^{-3}$.

\section{Calibration of the bias}\label{sec:two_filters}

In order to obtain accurate constraints on cosmological parameters,
the multiplicative bias for the {\it Euclid} cosmic shear analysis
needs to be less than $2\times 10^{-3}$
\citep{Masseyetal12,Cropper12}.  The results presented in the previous
section suggest, however, that the spatial variation of the SED will
lead to multiplicative biases in the ellipticity that exceed the
allowed range in the case of {\it Euclid}. Hence a way to mitigate the
problem of colour gradients is required.

In principle it should be possible to model the colour gradients (and
thus determine the bias) using resolved images of galaxies taken in
different filters. This approach was already suggested by
\cite{Voigtetal12}. Unfortunately, many factors influence the quality
of the results. The accuracy with which one can model the bias depends
on the properties of these images: signal-to-noise ratio, resolution,
pass-band characteristics, number of filters, properties and knowledge
of the PSF.

\cite{Voigtetal12} explored the possibility of a calibration sample,
covering the full range of properties of the source galaxies, for
which the bias can be determined. Following their work, we examine
this route in more detail, starting with the question what data are
required to determine the bias observationally.

\subsection{Approximated SED reconstruction}

To quantify the colour gradient bias, resolved observations in at
least two filters are needed. In this section we explore how well one
can reconstruct the spatial colour distribution of a galaxy when
observations in only two bands of that galaxy are available.  For each
of the narrower filters, we define the observed image:

\be \label{eq:interpole} I^{\rm obs}_{i}(\bx)=\int_{\Delta \lambda_i}
\lambda T_i(\lambda)S^0(\bx;\lambda) \ast P_i(\bx;\lambda) d\lambda\;, ~
i=1,2\,.  \ee

\noindent We use the observed image in each filter, $I_{i}^{\rm
  obs}(\bx)$, to derive the approximated intensity $S^{0, \rm
  approx}(\bx;\lambda)$ that we need to estimate the bias. We will for
the moment ignore the fact that a galaxy observed in the narrower 
filters has already been convolved by a PSF $P_i(\bx;\lambda)$ (i.e. we take it to be a
$\delta$-function). This will be addressed in the next
section. In addition we make the assumption that for each pixel the
SED can be interpolated linearly:

\be\label{eq:system} S^{0,\rm approx}(\bx;\lambda)=a({\bf x}) \lambda
+ b({\bf x}).  \ee

\noindent The coefficients $(a,b)$ can be determined for each
pixel by solving a linear system of equations: 

\be\label{eq:noconv} \int_{\Delta \lambda_i}\lambda T_i(\lambda) \Big [a({\bf
    x}) \lambda + b({\bf x}) \Big] d\lambda= I^{\rm obs}_{i}({\bf x})
\;, ~ i=1,2\,.  \ee

\noindent Once we obtain the approximated intensity $S^{0}_{\rm
  approx}(\bx;\lambda)$ we use it to estimate the bias induced by the
colour gradients in the same way we have done previously.

As shown below, {\it HST} observations represent the best available
sample to study the impact of colour gradients on the shear
estimation, which explains our choice of filters. We use the spatial
resolution and transmission $T_i(\lambda)$ for the F606W and F814W
filters (see left panel of Figure \ref{fig:filters}). Note that this
procedure can be applied to other sets of resolved observations taken
in two or more filters, although the accuracy of the results will
vary.  The use of more bands allows one to estimate the SED more
accurately, for example by using higher order polynomial
interpolations.

\begin{figure*} 
\psfig{figure=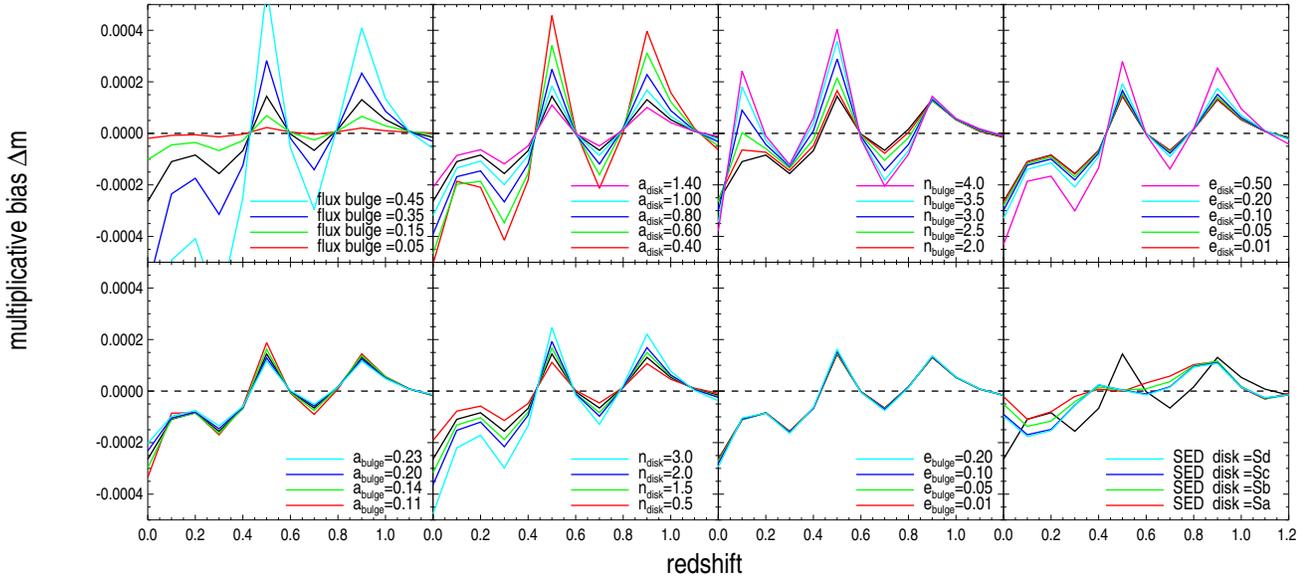,width=1.0\textwidth}
\caption{\label{fig:interpole} The difference between the predicted
  bias for an {\it Euclid}-like broad-band filter and the bias obtained by
  modelling spatially resolved observations in the {\it HST} F606W and
  F814W filters. The results are shown as a function of redshift for
  model galaxy B, varying the same parameters as was done in Figure
  \ref{fig:parameters}.}
\end{figure*} 

In the presence of noise, as is the case for real data, it will be
more challenging to reconstruct the spatial colour distribution by
interpolation. In the case of {\it Euclid} we do not expect this to be
a major issue as the galaxies used as calibration sample will have a
relatively high signal-to-noise ratio in typical {\it HST}
observations. Alternatively, one might consider ways to reduce the
impact of the noise, for example by using a shapelet decomposition
\citep{Refregier03,Kuijken06}.

Having fixed the properties of the two filters, we can compute the
difference $\Delta m$ between the bias for the {\it Euclid}
broad-band observations and the bias reconstructed using the
linear interpolation from F606W and F814W observations.  Figure
\ref{fig:interpole} shows the results for model galaxy B as a function
of redshift when we vary the parameters as was done for Figure
\ref{fig:parameters}. We find that we are able to recover the bias
with an accuracy of a few times $10^{-4}$.

We note that $\Delta m$ generally shows the same features and that its
amplitude is proportional to the input bias. This stems directly from
the fact that the linear interpolation cannot reproduce the Irr SED
accurately enough (see Figure \ref{fig:SED}).  The main reason for
this is associated with the presence of emission lines in the range
$400 \, {\rm nm} <\lambda< 600 \, {\rm nm}$, although this is not the
only reason.
Not surprisingly, the linear interpolation cannot reproduce any of the
SEDs perfectly over the large range in wavelength covered by the {\it
  Euclid} pass-band. The linear interpolation fails to capture
some of the features of the SEDs visible at $\lambda <500 \,{\rm nm}$
and the Balmer break at $400~ {\rm nm}$.

Hence the inaccuracy in modelling the SED at each position leads to a
residual bias which is a strong function of redshift. However, we note
that Figure \ref{fig:interpole} exacerbates the problem because the
local Irr SED is always the same and the linear interpolation fails
coherently. In practice, the linear interpolation will sometimes
overestimate and sometimes understimate the bias such that the average
bias as a function of redshift might still be estimated correctly and
the residuals uncorrelated. The SED also depends on the age of
the stellar population, the metallicity, the dust extinction, and the
velocity dispersion; all aspects we have neglected. For instance, the
bottom-right panel of Figure \ref{fig:interpole} shows that changing
the Irr SED used for the disk to any of the Sa-Sd spectra changes the
residual bias substantially.

Finally, it is possible to improve on these results by performing an
SED fit instead of a linear interpolation.  This might be feasible
especially in the case of low redshift galaxies for which multi-colour
observations are available and can be used to put tight constrains on
the local SED. This could significantly improve the accuracy on the
estimated bias at low redshift and help to reduce the residual bias
at higher redshifts.

%Unfortunately, the impact of this residual
%bias is hard to quantify and its characterisation should be done using
%real data, which is beyond the scope of this paper.

\subsection{Effect of the native PSF}

Since the images we use to evaluate the colour gradient-induced bias
have been convolved with a PSF themselves, the procedure used to
retrieve the local SED is more complicated in practice. Accounting for
the PSF, Equation (\ref{eq:interpole}) changes to

\be\label{eq:deconvolution} \int_{\Delta \lambda_i} \lambda T_i(\lambda) [a(\bx)
  \lambda + b(\bx)] \ast P_i(\bx;\lambda) d\lambda = I^{\rm
  obs}_{i}(\bx) \;, ~ i=1,2\,.  \ee

\noindent In Fourier space we obtain a linear system of equations
which we can solve to obtain $a(\bk)$ and $b(\bk)$, the Fourier
transformed maps of the linear coefficients to approximate the local
SED. Solving a linear system in Fourier space corresponds to
performing a deconvolution. This will therefore cause loss of
information, as we cannot reconstruct scales that are smaller than the
PSF. Note that this system of equations can be expanded to more
filters, and higher order interpolations. The filters might also have
different PSFs.

%It is interesting to quantify how the SED reconstruction will perform
%in practice when one uses observations with different PSFs, different
%spatial resolutions, and different noise properties. We will address
%these issues in future work. We will instead continue to focus our
%study on the {\it HST} case.

\section{Euclid bias modelling with {\it HST} filters}\label{sec:HST}

In the previous section we have shown that it is possible to quantify
the effect of colour gradients using resolved observations in at least
two filters. A complication is that one needs to deconvolve the
galaxies to account for the native PSF of the narrower filters. This
will reduce the overall accuracy of the measurement of the colour
gradient bias for two reasons. First of all, solving the system of
equations (\ref{eq:deconvolution}) for the local SED is equivalent to
a deconvolution, implying that there is an upper limit to the spatial
frequencies we can recover. Secondly, the native PSF we need to
correct for is not perfectly known.

\begin{figure}
\begin{tabular}{|@{}l@{}|@{}l@{}|} 
\psfig{figure=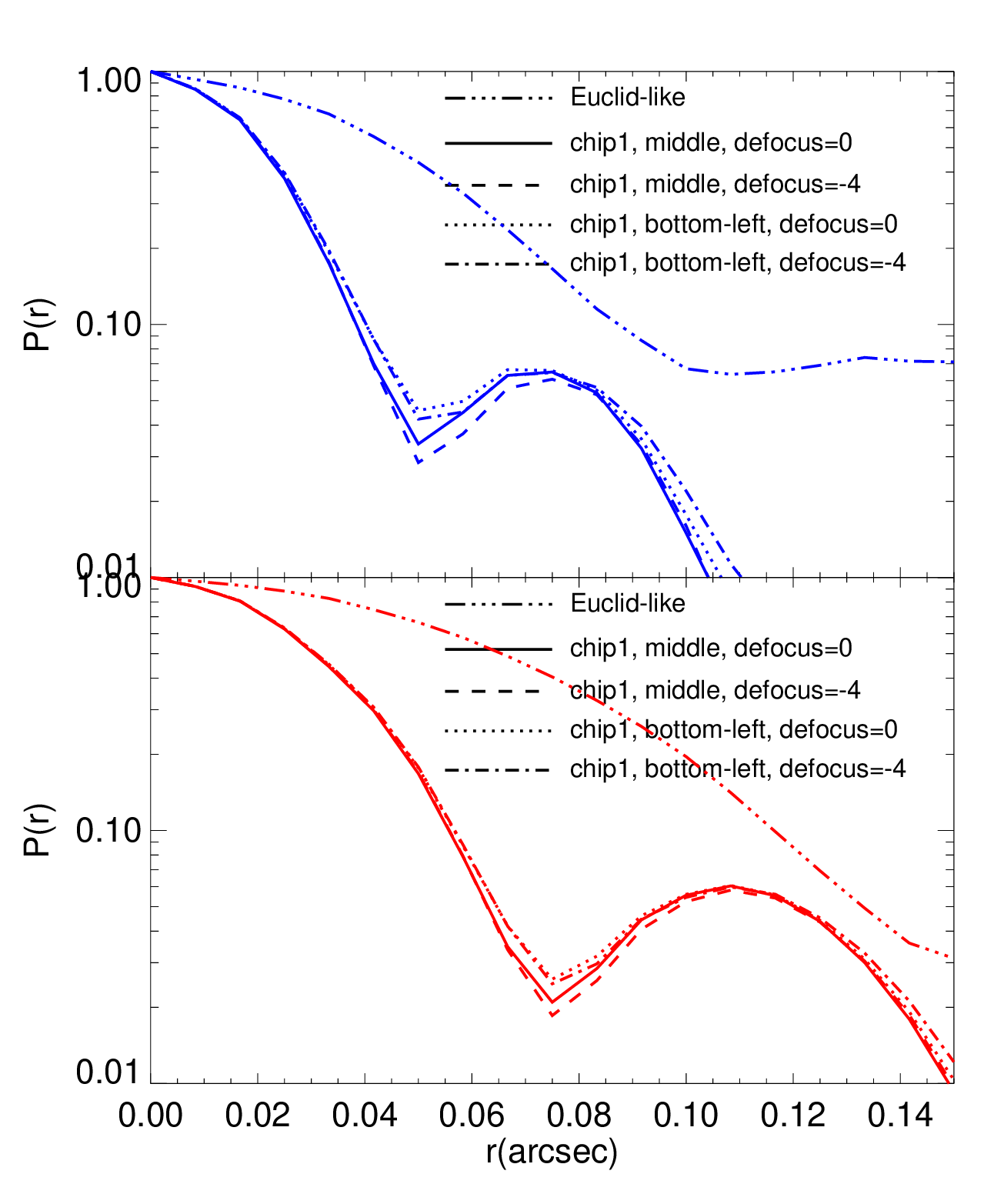,width=0.45\textwidth}
\end{tabular}
\caption{\label{fig:HST_PSF}Top panel: Comparison of the {\it Euclid}
  PSF profile at $\lambda=550~ {\rm nm}$ (solid-dotted line) with the
  ACS (Wide Field Channel)-{\it HST} PSF profiles at the same
  wavelength. For the {\it HST} PSF we show the different profiles as
  a function of position and defocus values expressed in $\mu m$.
  Bottom panel: the same as top panel but for $\lambda=800~ {\rm nm}$.
}
\end{figure}

Since deconvolution algorithms have intrinsic limitations which depend
on the size of the PSF, a small PSF is always preferable.  Amongst
currently available data, {\it HST} observations are therefore the most
suitable to model colour gradients. As can be seen in Figure
\ref{fig:HST_PSF}, the characteristic size of the {\it Euclid} PSF is
twice the {\it HST} PSF. This stems directly from the fact that the
diameter of the {\it Euclid} mirror $D=1.2\, {\rm m}$ is about half of
the size of the {\it HST} mirror, $D=2.5\, {\rm m}$. Despite its small
size, there is a loss in accuracy caused by the {\it HST}
PSF. Furthermore, the {\it HST} PSF does vary as a function of position
and time, which can be modelled with finite accuracy. The impact
of both complications are quantified below.

We generate images of the {\it HST} PSF at different positions and for
various focus configurations using the Tiny-Tim software
\citep{KrHoSt11}.  The PSF images $P(\bx;\lambda)$ have a pixel size
of $0.05 ~ {\rm arcsec}$ and are sampled as a function of wavelength
with a step $\delta \lambda=50~ {\rm nm}$. When convolving the images
of the source  $I^0(\bk;\lambda)$ with the {\it HST} PSF, we
approximate the PSF to the closest one in $\lambda$ instead of
interpolating between the various PSF images.

The top panel of Figure~\ref{fig:HST_PSF} shows our nominal {\it
  HST} PSF profile, which is the one in the middle of the first ACS
chip, without defocus, for $\lambda=550~ {\rm nm}$. We compare this
profile to those obtained by changing the position across
the camera and the focus. In the bottom panel we show the same {\it
  HST} PSF profiles but for $\lambda=800~ {\rm nm}$.  Note that we
ignore the camera distortion, as it does not alter colour
gradients and can be corrected for before modelling the bias.

Both focus and position changes from the nominal configuration
increase the half-light radius of the PSF, but in different ways. The
change of the focus (upper panel of Figure \ref{fig:HST_PSF}) does not
affect the core of the PSF but it affects the wings, which become
larger (effectively the PSF profile is slightly suppressed at small
scales and boosted at larger scales). Changing the position affects
both the core and the wings of the PSF. This is the result of a change
in the diffusion coefficients which increases the overall
characteristic size of the PSF. In both cases the effect is very small
and we do not expect the final results to depend significantly on the
position nor on the focus.

\subsection{PSF deconvolution}

\begin{figure*}
\begin{tabular}{|@{}l@{}|@{}l@{}|} 
\psfig{figure=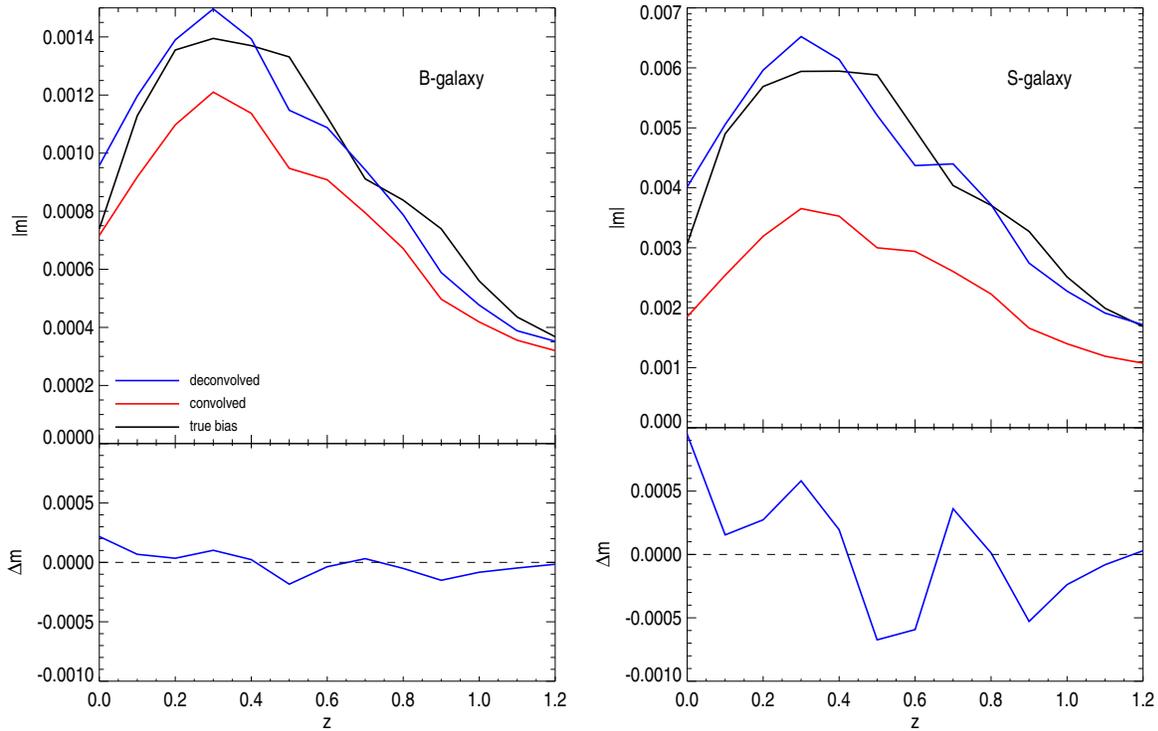,width=0.9\textwidth}
\end{tabular}
\caption{\label{fig:deconvolved} Left figure, top panel: Comparison of
  the true bias for model galaxy B (black solid line) with the bias
  using a linear interpolation of the SED using the F606W and F814W
  filters, but ignoring the effect of the {\it HST} PSF. The red solid
  line indicates the resulting bias for the reference {\it HST} PSF.
  The blue solid line shows the bias estimate when we do account for
  the {\it HST} PSF. The bottom panel shows the difference between the
  true bias and its estimate accounting for the {\it HST} PSF. The
  right panels show the same as the left panels but for the reference
  galaxy S.  }
\end{figure*}

We use the PSFs generated with Tiny-Tim for the nominal focus and
position to create images for model galaxies B and S. We estimate the
bias without accounting for the PSF and present the results in
Figure~\ref{fig:deconvolved} (red lines). Comparison with the
true bias (black lines) shows that ignoring the PSF leads to an
underestimate of the bias because the native PSF blurs the colour
gradients. As expected, the effect is larger when the galaxy is
smaller (right panel).

If we instead use Equation (\ref{eq:deconvolution}) to reconstruct the
approximated intensity profiles we obtain the blue solid lines in
Figure~\ref{fig:deconvolved}. Hence, when the convolution by the {\it
  HST} PSF is accounted for, we recover the original bias quite well,
as shown in the bottom panels where we plot the difference $\Delta m$
between the original bias and the estimated one. For the galaxy B we
are able to recover the bias within an accuracy of $2 \times 10^{-4}$;
as expected, the performance is worse for the S galaxy. This
difference includes the error made by approximating the local SED with
a linear function and it is interesting to note that the residuals for
both galaxies are similar to the one estimated in Section
\ref{sec:two_filters}, where we ignored the PSF of the narrower band
data. This suggests that the linear interpolation of the SED is still
the main limitation and that {\it HST} data are suitable to model the
bias induced by the presence of colour gradients. As a caveat we note
that the deconvolution step will be more complicated in practice due
to the presence of noise in real data.

\subsection{Imperfect PSF knowledge}

The {\it HST} PSF cannot be modelled perfectly because it varies with
time. The resulting error in the PSF model will lead to an error in
the estimate of the colour gradient bias. We estimate the impact of
this by examining the variation in the bias for a range of {\it HST}
PSFs. Under the assumption that the deconvolution and the SED
interpolation work perfectly, the loss of accuracy is given by the
dispersion between the bias estimates for the various PSFs.  This
allows us to estimate an upper limit to the accuracy due to the errors
in the {\it HST} PSF model.

\begin{figure*}
\begin{tabular}{|@{}l@{}|@{}l@{}|} 
\psfig{figure=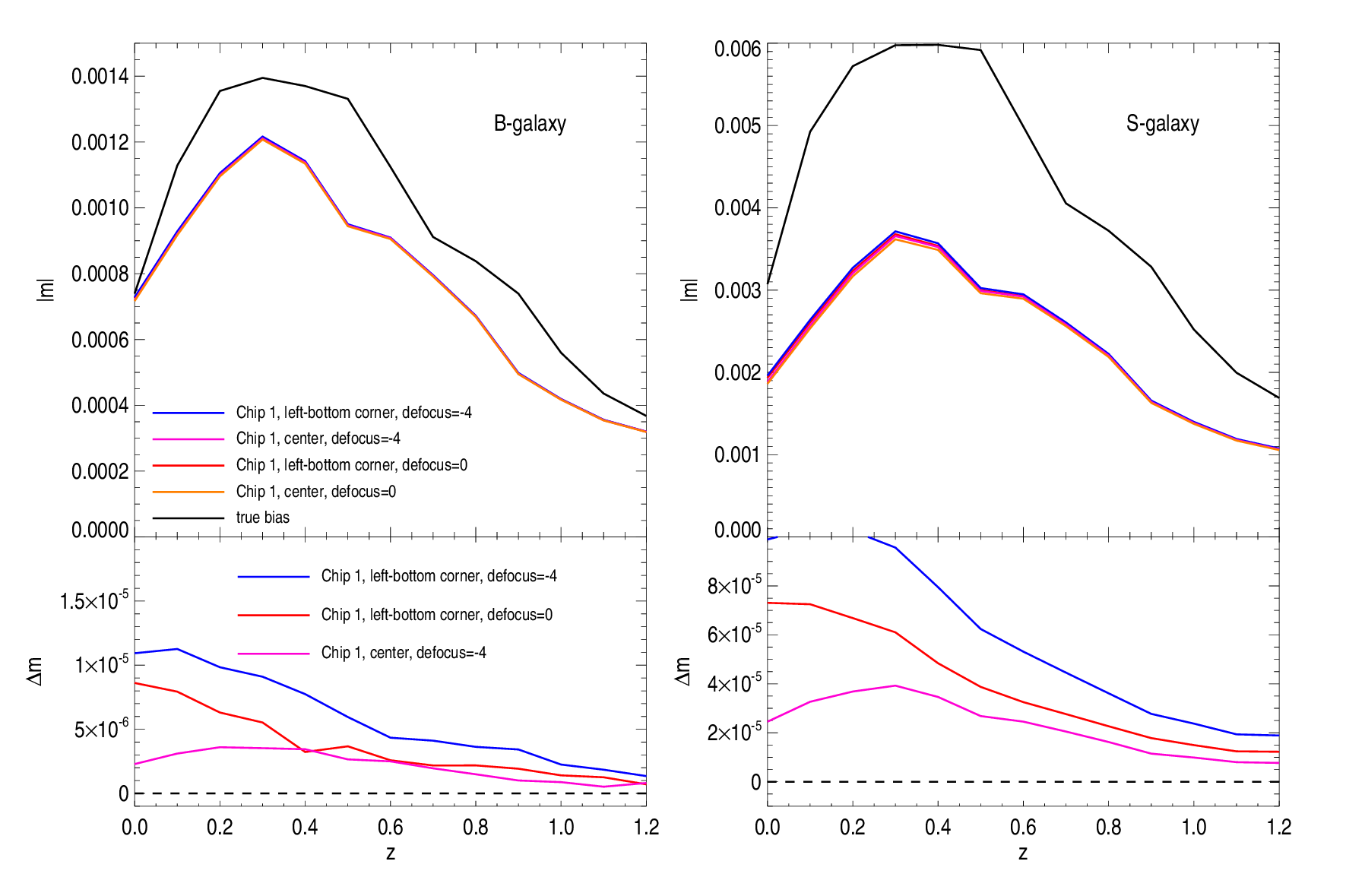,width=0.9\textwidth}
\end{tabular}
\caption{\label{fig:HST_convolved} Left figure, top panel: Comparison
  of the true bias for model galaxy B (black solid line) with the bias
  using a linear interpolation of the SED using the F606W and F814W
  filters, but ignoring the convolution by the {\it HST} PSF. The
  orange solid line indicates the resulting bias for the reference
  {\it HST} PSF. The other lines indicate how the bias changes as a
  function of the position and defocus value expressed in $\mu m$. The
  bottom panel shows the difference between the value of the
  multiplicative bias obtained for a galaxy observed using the
  reference {\it HST} PSF, and the bias when the position of the
  galaxy is changed and/or a defocus is present. The right panels show
  the same results, but for the S galaxy.}
\end{figure*} 

To do so, we evaluate the bias for slightly different PSFs but without
deconvolving the PSF. We generate PSFs using Tiny Tim for various
focus configurations and camera positions and measure the resulting
colour gradient bias using Equation (\ref{eq:noconv}). The top panels
of Figure \ref{fig:HST_convolved} show the amplitude of the bias for a
few positions and defocus values. For reference we also show the true
bias (black solid line). The difference between the bias for the
reference {\it HST} PSF and the bias obtained when the defocus and the
position are changed is presented in the bottom panels. Note that the
differences are always smaller that $10^{-4}$ and always positive.
This is because the changes in position and focus both increase the
PSF size (and thus the absolute value of the bias).  As expected, the
impact of variations of the {\it HST} PSF are larger for galaxy S. For
galaxy B, the changes in the bias values are about $10^{-5}$, whereas
for the galaxy S they are about $10^{-4}$ in the worst case.

Were it possible to perform a perfect deconvolution, ignoring PSF
variations as a function of position and focus would lead to an  extra error term. This error can be roughly estimated as the product of the dispersion from
the various focus and position configurations, multiplied by the ratio
between the true and recovered bias (black and orange lines,
respectively). In practice the PSF as a function of position and focus
can be determined reasonably well (see for example
\citealt{Rhodesetal07,Schrabbacketal10}) and the error is expected to
be significantly smaller than what is shown in Figure
\ref{fig:HST_convolved}. Hence, the limited accuracy of the model for
the {\it HST} PSF is not an important source of error.

% From this study it emerges that {\it HST} data are well suited to
% calibrate the colour gradient induced bias. The PSF size and its
% knowledge are such that it is possible to use {\it HST} data to model
% the colour gradient induced bias affecting {\it Euclid} shear
% estimates. The main limitation to model this bias is given by the
% poor approximation of the local SED.  It is possible to improve the
% accuracy either using resolved observations in more bands or using a
% better parametrisation of the local SED.

\section{Calibration of the  bias}\label{sec:sample}

\citet{Voigtetal12} proposed to use {\it HST} observations of galaxies
to determine the mean bias as a function of galaxy properties. The
precision with which this can be done depends on the intrinsic
variation in galaxy properties, which in turn drives the size of the
sample of galaxies that is needed. In this section we examine whether
the {\it HST} archive contains a sufficient number of resolved
galaxies observed in at least two bands to model the bias with the
precision required for {\it Euclid}'s science objectives.

Throughout this paper we have made realistic but rather conservative
assumptions in order to estimate the bias using simulated bulge plus
disk galaxies that by design showed significant colour gradients.  The
actual amplitude of the bias, however, needs to be derived using
actual observations. A preliminary analysis based on 12,000 galaxies
in the Extended Groth Strip (EGS, \citealt{Daetal07}) suggests an
average value of $\ave{m} \approx 3 \times 10^{-3}$ in the worst
cases, with an uncertainty $\sigma_{\ave{m}}$ of a few times $10^{-4}$
computed in redshift bins each containing about 500 galaxies (Huang et
al. in prep.). 

The average bias is larger than the $3 \times 10^{-4}$ measured by
\citet{Voigtetal12} using their full dataset comprising $\sim 700$
galaxies. A source for this difference is our use in this paper,  and in the preliminary analysis of EGS data, of a PSF that has a larger size and a stronger $\lambda$ dependence than the one used by \citet{Voigtetal12} which is closer to  {\it Euclid}'s one. Additional differences might arise from the fact that we
use different shear estimates. Furthermore, as pointed out by
\citet{Voigtetal12}, their selection of galaxies may not be very
representative. For instance, the EGS dataset may contain a fraction
of galaxies with red bulges larger than the ones in the
\cite{Simardetal02} catalogue; as shown in Figure 6 of
\citet{Voigtetal12}, selecting galaxies with a redder bulge enhances
the bias to $10^{-3}$. \citet{Voigtetal12} estimate a dispersion in the bias of $ \sim 7 \times 10^{-3}$ for the whole sample, which suggests that about
a thousand\footnote{\cite{Voigtetal12} quote a number of galaxies that
  is ten times larger to account for possible limitations in the
  accuracy of their analysis, which is based on a small sample of
  galaxies with a bulge and disk decomposition.}  galaxies per
redshift bin are needed to obtain $\sigma_{\ave{m}} \simeq 2 \times
10^{-4}$. The error on the average we obtain from the preliminary EGS
analysis agrees with these values. As shown in Table \ref{tab:2}, the
{\it HST} archive contains enough galaxies to reach this precision.

\subsection{Residual bias correlations}

The bias depends on the intrinsic properties of the galaxies, which in
turn depend on environment. The colour gradient bias will therefore
vary spatially. A simple correction using the average bias for each
tomographic bin may therefore not be sufficient, and still lead
to a spurious signal in the two-point cosmic shear statistics.

We assess the residual bias on the correlation function as follows. We
compute the two-point ellipticity correlation function $\xi_+(\theta)$
selecting sources belonging to the same redshift bin:

\be\label{eq:xi}
\xi_+(\theta)=\ave{\gamma_t(\thetag^\prime)\gamma_t(\thetag+\thetag^\prime)}+\ave{\gamma_r(\thetag^\prime)\gamma_r(\thetag+\thetag^\prime)}\,,
\ee

\noindent where the ensemble average is meant as an average over all
pairs with distance $\theta$ and $\gamma_t$ and $\gamma_r$ indicate
the tangential and the $45$-degree rotated components of the
estimated shear projected along the line connecting the pair of
galaxies \citep[see for example][]{BaSc01}. We define
$\Delta\xi(\theta)$, the relative bias in the correlation function:

\be \Delta \xi(\theta)=\frac{\xi^{\rm obs}_{+}(\theta)}{\xi_{+}(\theta)}-1\,.
\ee

Based on the breakdown presented in \cite{Cropper12} we assume that
{\it Euclid}'s science objectives can be achieved if the value
$|\Delta \xi(\theta)|$ is smaller than $5 \times 10^{-4}$ for all
angular scales $\theta > 6~ {\rm arcmin}$. This limit allows us to
estimate the required minimum size of the calibration sample that is
used to determine the average colour gradient bias as a function of
galaxy properties and redshift.

Since we lack observational results we use simulations to
conservatively estimate the impact of the spatial variation in the
colour gradient bias. In the following we use model galaxy S, which
has a bias of a few times $10^{-3}$ and an observed FWHM of about
$1.4\,{\rm FWHM_{PSF}}$, typical of a galaxy in the {\it Euclid}
survey.

To simulate the properties of galaxies in the {\it Euclid} survey and
the calibration sample we use ray tracing simulations produced using
the Millennium Simulation \citep{Hietal09,Springel05}.  The ray
tracing simulations comprise $32$ lines of sight, each covering
$16~{\rm deg^2}$.  For each galaxy we know the shear, the
redshift and magnitudes in various SDSS bands. To simulate the depth
of the {\it Euclid} catalogue we add the $i$, $r$ and a third of the
$z$ fluxes of the SDSS bands and select galaxies brighter than $m_{\rm
  riz}=24.5$. The resulting density is $32~ {\rm gal/arcmin^2}$, in
agreement with expectations. We use the SDSS $r-i$ values to define
the colour of each galaxy.

\begin{figure*}
\begin{tabular}{|@{}l@{}|@{}l@{}|} 
\psfig{figure=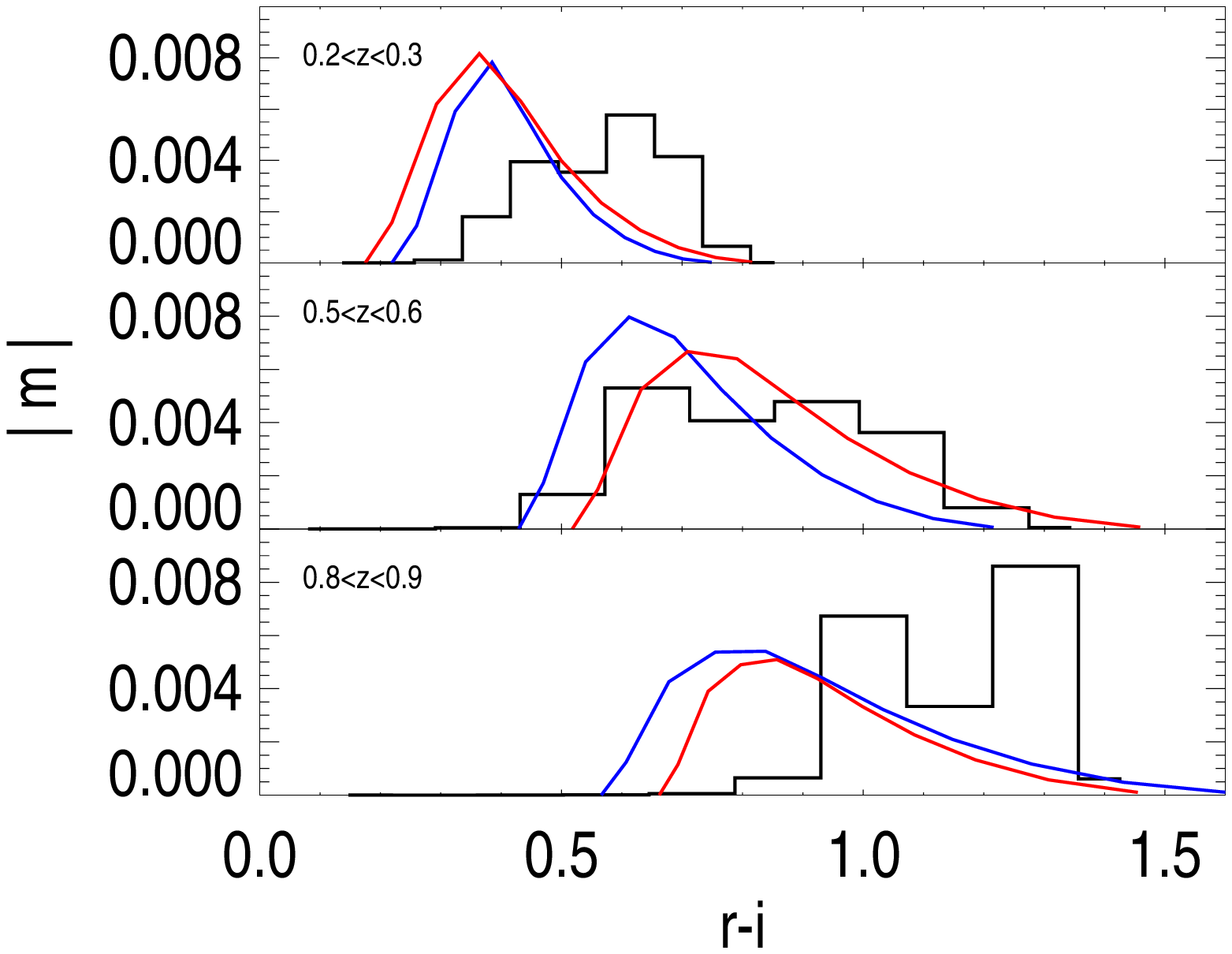,width=0.5\textwidth}&\psfig{figure=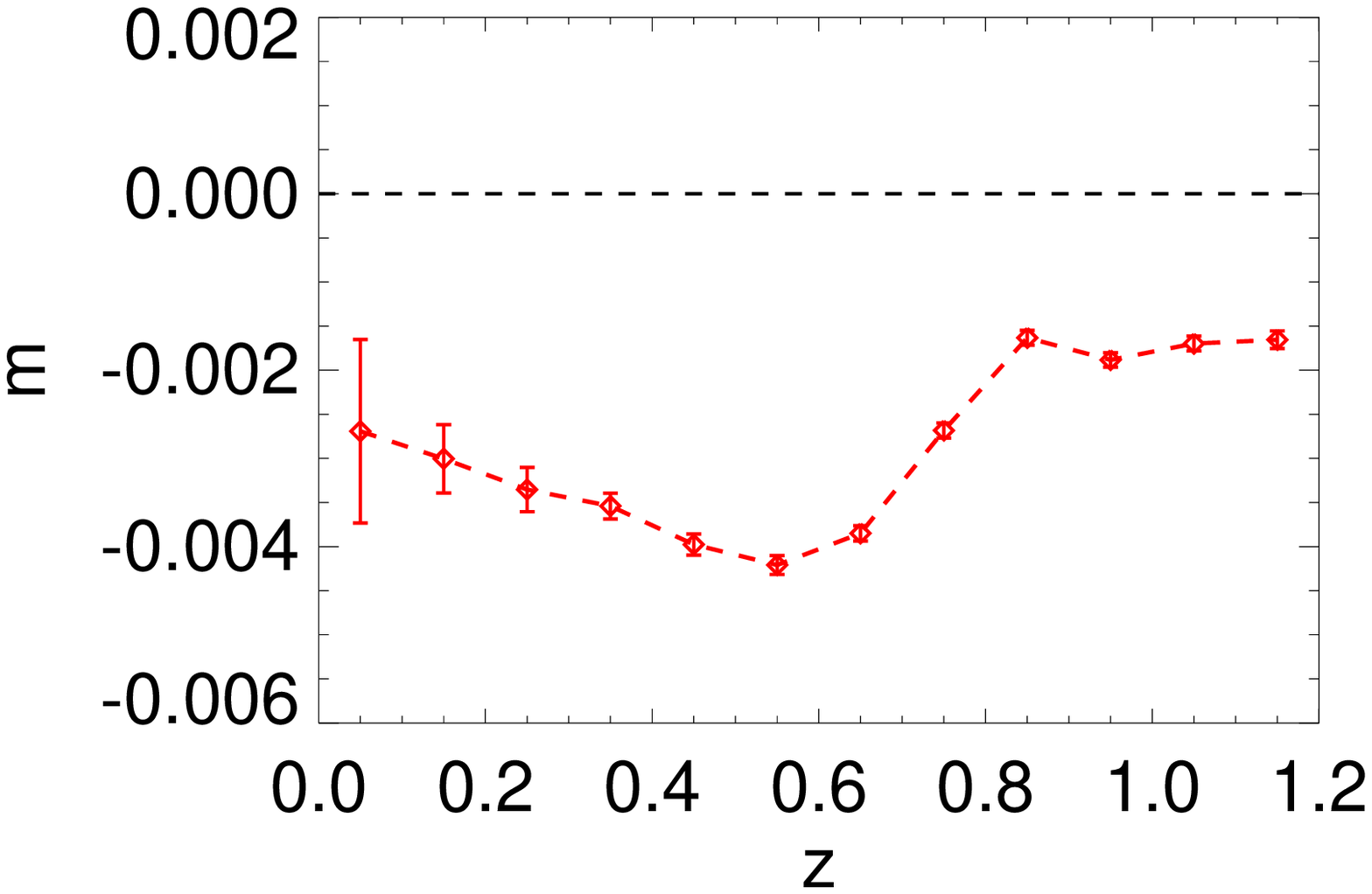,width=0.5\textwidth}\\
\end{tabular}
\caption{\label{fig:bias} Left panel: Distribution of `observed' SDSS
  $r-i$ colours of the simulated galaxies in three redshift bins. For
  each redshift bin, the blue (red) solid line represents the absolute
  value of multiplicative bias predicted by our model (see text) as a
  function of observed colour for the sources with the lowest
  (highest) redshift included in that bin. Right panel: Average bias
  $\ave{m}$ and its error $\sigma(\ave{m})$ using the calibration
  sample of $\sim 62,000$ galaxies to which we assign a bias as a
  function of colour and redshift as described in Section
  \ref{sec:sample}. }
\end{figure*} 

\subsection{Size of the calibration sample} 

We searched the {\it HST} archive for ACS observations in at least two
(suitable) filters. The results are listed in Table~\ref{tab:2}, where
we note that we also include scheduled observations. The sample
contains approximately $1536 \, {\rm arcmin^2}$ of F606W and F814W
data including EGS \citep{Daetal07} and the overlap of CANDLES
\citep{Gretal11} with either COSMOS \citep{Scetal07} and GOODS
\citep{Gietal04}. We have also included the area of GEMS
\citep{Rietal04} data that does not overlap with either GOODS or
CANDLES. For this part of the GEMS data we have F850LP observations which
are shallower and slightly redder than the F814W data; these can be
also used to calibrate colour gradients, but the lower signal-to-noise
is expected to reduce the accuracy. This results in a total area of
$2056~{\rm arcmin^2}$ containing about $62,000$ galaxies for which we
can determine the colour gradient bias observationally.

The data listed in Table~\ref{tab:2} were obtained as part of several
surveys. Including single ACS pointings as well would increase the
area by another $\sim 1000\, {\rm arcmin^2}$. Finally, dedicated deep
{\it Euclid} observations of the area covered by STAGES
\citep{Gretal09} and COSMOS would increase the sample by more than a
factor of three. The full benefit of the latter observations requires
more study because the {\it Euclid} PSF size is larger and pass-band
broader. In the following we therefore conservatively assume that we
can measure the colour gradient bias using {\it HST} observations of
$62,000$ galaxies.

\begin{table}
\begin{tabular}{lll}
Name   & Area (${\rm arcmin^2}$) & Number of galaxies \\
\hline
EGS &    650 & 18000  \\              
CANDELS/UDS &   198 & 6000\\                  
CANDELS/COSMOS &   $198^\star$ & 6000  \\                   
CANDELS/GOODS-CDFS & $300^\star$  & 9000 \\                   
CANDELS/GOODS-NORTH & $190^\star$ & 5500\\                    
GEMS/CANDLES+GOODS & $520$ (F850LP) & 15500 \\
\hline
total &  $2056 $ & 62000\\
\end{tabular}

\caption{\label{tab:2} Size of the {\it HST} data sample observed in
  both F606W and F814W/F850LP bands. The entries marked with $\star$
  are not yet (fully) available, but will be soon. All observations
  quoted in the table are deeper than {\it Euclid}. The number of
  galaxies has been computed assuming ${\rm F814W} < 24.5 $ which
  matches the number density of $30 {\rm \,gal/arcmin^2}$ expected for
  {\it Euclid}. Note that the F850LP observations are shallower but
  they can still be used with in combination with the F606W data to
  obtain an estimate for the local SED.}
\end{table}

\subsection{Bias model}

To each galaxy in the simulation we assign a bias $m$, which is a
function of colour and redshift. Including the colour dependence is
important to capture the fact that galaxy colour depends on
environment. The value for the bias is obtained by taking the S galaxy
and varying the fraction of light in the bulge as we did in the
top-left panel of Figure \ref{fig:parameters}. Note that increasing
the fraction of light in the bulge without changing its size results
in redder galaxies with an unrealistically small but very bright
bulge. To avoid this, we increase the size of the bulge accordingly,
thus accounting for the increase in its flux. The galaxy S has a bulge
size of $r_h=0.09\, {\rm arcsec}$ which contains 25\% of the light; we
set $r_h$ of the bulge to that of the disk ($r_h =0.59\, {\rm
  arcsec}$) when the flux of the bulge is $100\%$ (note that in this
case we have an elliptical galaxy with no colour gradients) and create
all other cases using a linear relation between flux and size.

Once we have created this set of model galaxies, we derive their
colours by integrating the flux over the SDSS $i$ and $r$ filters.
The resulting colours are generally too blue compared to the galaxies
in the Millennium Simulation. These differences are likely caused by
differences in the adopted SEDs. The change in colour with redshift
for our model early-type galaxies is the same as that of the red
Millennium galaxies. Thus, we correct for this offset in colour by
shifting them slightly to match the results from the Millennium
Simulation.

%For instance, it is known that single stellar population synthesis
%models such as the one we use here cannot reproduce the observed
%colours \citep{Marastonetal09}.

\begin{figure*}
\begin{tabular}{|@{}l@{}|@{}l@{}|} 
\psfig{figure=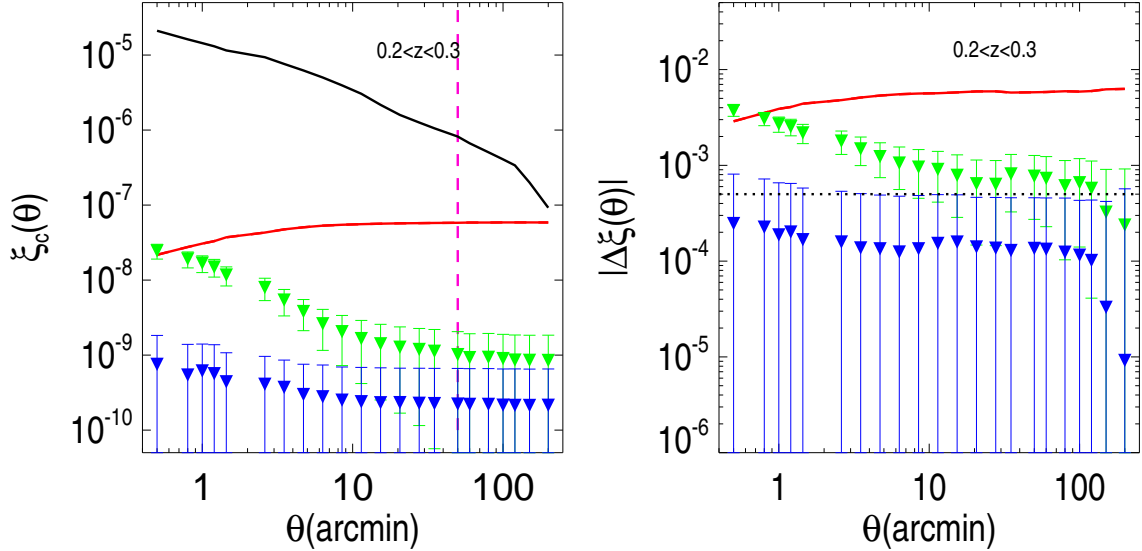,width=0.9\textwidth}
\end{tabular}
\caption{\label{fig:m_and_c} Left panel: The spurious contribution to
  the ellipticity correlation function due to colour gradient bias
  $\xi_c(\theta)$ (red solid line) compared to the cosmic shear signal
  $\xi_+(\theta)$ for a tomographic bin with $0.2<z<0.3$ (black solid
  line). The magenta dashed line shows the angular scale corresponding to
  the {\it Euclid} field-of-view. The green triangles indicate the
  spurious signal if the bias correction is a function of redshift
  only. If the correction is a function of both colour and redshift
  the blue triangles are obtained. Right panel: The absolute value of
  the relative change in the ellipticity correlation function if
  $\xi_c=0$. The colours and symbols are the same as for the left
  panel. The dotted black line indicates the maximum allowed value for
  $|\Delta \xi|$ for {\it Euclid} to meet the requirements.  The errors represent the standard deviation
  from $32$ independent realisations of the calibration sample (see
  text).}
\end{figure*}

The left panel of Figure \ref{fig:bias} shows the colour distribution
of the simulated {\it Euclid}-like galaxies for three redshift bins.
For each redshift bin we also show the bias as a function of the
observed colour. The blue (red) lines indicate the bias for a galaxy
at the lower (upper) limit of the respective redshift bin.  In all
cases the bias reaches a maximum value when the bulge contains about
20-30\% of the light.

The bias $m$ assigned to each galaxy in the Millennium Simulation is computed
by interpolating the value of the multiplicative bias $m$ as a
function of the $r-i$ colour at the redshift of the galaxy. The
average bias as a function of redshift is presented in the right panel
of Figure~\ref{fig:bias}. Figure~\ref{fig:bias} shows that $-4\times
10^{-3}<m<-2\times 10^{-3}$, similar to what is found from the
preliminary EGS analysis. The dispersion of the simulated bias,
however, is about $1\times 10^{-3}$, smaller than that found
from the EGS analysis. To ensure that our estimates remain
conservative, we increase the variance of $m$ accordingly: we include
an additional spread in the bias sampled by a Gaussian of width
$\sigma=6\times 10^{-3}$.

This results in a simulated data set where the bias depends on
redshift and colour with an extra variance to account for variations
as a function of other parameters (such as ellipticity and size).
These estimates were used to compute the errors on the mean bias as a
function of redshift presented in Figure~\ref{fig:bias}, where we
assume a calibration sample of $62,000$ galaxies ($40,000$ of which
have $z<1.2$).

\subsection{ Effect on the ellipticity correlation function}

So far we have limited the discussion to the impact of colour
gradients on the multiplicative bias $m$. As the bias is related to
the correction for the PSF, we expect the additive bias $c$ to be
affected as well. In this more general case, the shear estimate for
the $i$-component is given by:
$$\tilde \gamma_i=(1+m) \gamma_i+c.$$

\noindent Inserting this expression into Equation (\ref{eq:xi}) we find
that the measured correlation function can be expressed as:

\be \label{eq:corr}
\xi^{\rm obs}_{+}(\theta)=\xi_+(\theta) \big[1+2 \ave{m} + \xi_m(\theta)\big]+ \xi_c(\theta)\,,
\ee

\noindent where $\xi_m(\theta)$ and $\xi_c(\theta)$ quantify the
correlations of $m$ and $c$, respectively, as a function of angular
separation. The relative bias in the correlation function is thus
given by:

\be \label{eq:corr}
\Delta \xi (\theta)=[2 \ave{m} + \xi_m(\theta)\big]+ \frac{\xi_c(\theta)}{\xi_+(\theta)}\,.
\ee

\noindent \cite{Masseyetal12} derived expressions for the various sources of
multiplicative and additive bias and showed that the contributions to
$m$ and $c$ caused by colour gradients are related through:

\be c=m \frac{ e_{{\rm PSF},i}}{P_\gamma P_{e_ {\rm PSF}}}, \ee

\noindent where $e_{{\rm PSF},i}$ is the $i$-component for the PSF
ellipticity. Following \cite{Masseyetal12}, we assume $P_{e_{\rm PSF}}
= 1$. Hence we have $\xi_c(\theta)=\xi_m(\theta)\xi_{+,{\rm
    PSF}}(\theta)/P_\gamma^2$, where $\xi_{+,{\rm PSF}}(\theta)$ is
the ellipticity correlation function of the PSF anisotropy.  Because
our definition of ellipticity by Equation~(\ref{eq:defellipticity})
differs from the one used by \cite{Masseyetal12}, we have
$P_\gamma\simeq 1 -e^2\simeq 0.86$ whereas we adopt $e_{\rm PSF} =
7\%$ for the PSF ellipticity\footnote{This corresponds to the maximum
  value for the polarisation of ~$15\%$ allowed for the {\it Euclid}
  PSF \citep{redbook,Cropper12}.}, which is an extremely pessimistic
scenario.

Although $c$ is smaller than $m$, Equation~(\ref{eq:corr}) shows that
the additive bias can become dominant on large scales if the PSF
ellipticity correlation function does not vanish sufficiently quickly.
In our case the PSF orientation is constant across the field, which
implies that for any given value of $c$ there will be a scale for
which $\xi_c(\theta) > \xi_+(\theta)$, unless $\xi_m(\theta)$ vanishes
faster than $\xi_+(\theta)$. This scale is smaller for low-redshift
sources for which the cosmic shear signal is smaller. 

This large-scale signal provides a way to estimate the amplitude of
the colour-gradient induced bias directly from the {\it Euclid}
dataset by measuring the correlation between the ellipticity of stars
and the PSF-corrected ellipticity of galaxies (i.e. the shear).  This
correlation is commonly used as a diagnostic to ensure that the PSF
anisotropy is well corrected \citep[e.g.][]{Heymans12}.  For this
reason one would rather not use this residual signal to correct for
colour gradients. However, in the case of colour gradients we do not
need to correlate galaxies and stars in the same field: the colour
gradient bias will be proportional to the PSF anisotropy, so that
residuals of independent fields will be correlated. Note that in the
absence of colour gradients these correlations are supposed to vanish
if the correction for PSF anisotropy is accurate on average.

\begin{table*}
\begin{tabular} {lllllll}
& \multicolumn{3}{l}{redshift only} & \multicolumn{3}{l}{redshift and colour}\\
 & $0.2<z<0.3$ & $0.5<z<0.6$& $0.8<z<0.9$ & $0.2<z<0.3$ & $0.5<z<0.6$& $0.8<z<0.9$  \\
\hline\
{ $\theta < 240~ {\rm arcmin}$ } & $34.5 \pm 14.6$ & $ 1.0 \pm 11.8 $ &  $ 2.1 \pm  5.4 $ & $ 5.7 \pm 4.4$ &  $-5.1 \pm  11.3$ &  $-2.2\pm 2.2$ \\
{ $\theta > 6~ {\rm arcmin}$ } &  $31.3 \pm 16.1$ &  $-1.4\pm 13.4$ &  $ 1.3 \pm 6.5$ & $7.4\pm 4.8$ & $-6.3 \pm 14.2$ & $-2.8 \pm 2.6$ \\
{ $6<\theta < 50~{\rm arcmin}$ }  &  $28.0\pm 8.9$ &  $-1.7 \pm 1.5$ &  $1.2 \pm 0.4 $ &  $ 6.1 \pm 1.6$ & $-3.6 \pm 3.3$ &  $-1.3\pm 0.3$ \\

\end{tabular}
\caption{\label{tab:tabxi} Average relative bias $\Delta \xi (\theta)$
  in units of $10^{-4}$ for the three redshift bins presented in
  Figure \ref{fig:bias_corr}.  Columns $2-4$ list results if the bias
  is modelled as a function of redshift only; Columns $5-7$ list
  result when the bias is modelled as a function of both redshift and
  colour.}
\end{table*}

\begin{figure*}
\begin{tabular}{|@{}l@{}|@{}l@{}|} 
\psfig{figure=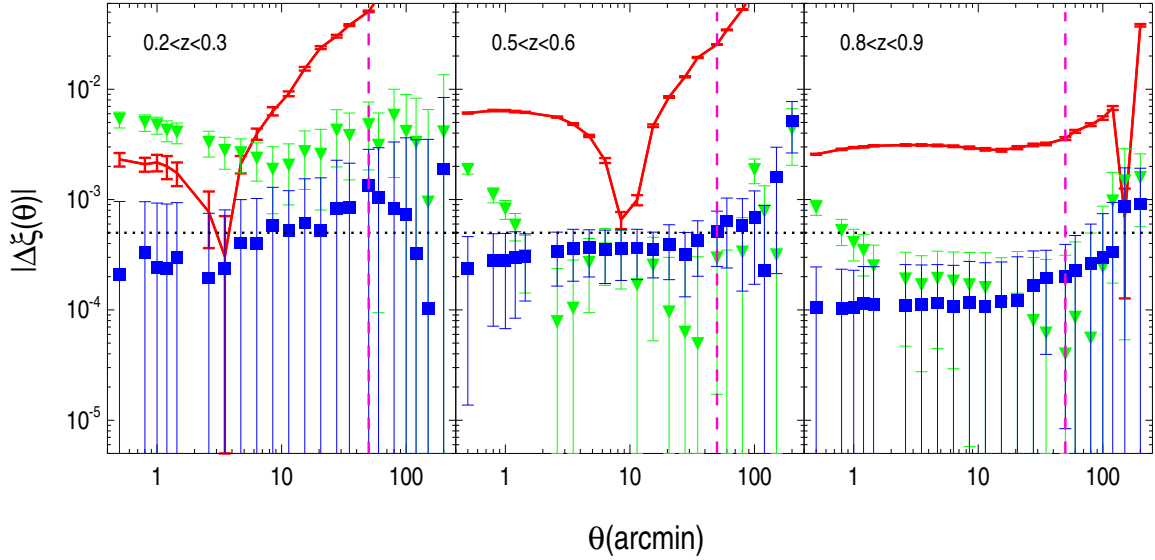,width=0.9\textwidth}
\end{tabular}
\caption{\label{fig:bias_corr} Absolute value of the bias $\Delta \xi
  (\theta)$ as a function of scale for three redshift bins. The dotted
  line shows the maximal residual bias allowed to meet {\it Euclid}'s
  science requirements. For reference, the vertical dashed line
  indicates the size of the {\it Euclid} field-of-view. The red line
  shows the relative bias in the ellipticity correlation function when
  colour gradients are ignored. In this case the multiplicative bias
  lowers the signal (since $m<0$), but the additive bias leads to an
  increase in signal. The green triangles show the relative bias when
  the correction is a function of redshift only. The results improve
  when the correction is a function of colour and redshift (blue
  squares). The errors represent the standard deviation from $32$
  independent realisations of the calibration sample.  The mean values of the residual bias and their errors are listed in Table \ref{tab:tabxi}.}
\end{figure*}

The calibration sample is used to obtain an estimate for $\ave{m}$,
the average value of the bias for a particular selection of sources,
which is used to correct the shear signal for these galaxies. Hence
the relevant quantity becomes $\delta m$, the residual bias for each
galaxy after correction. After removing the average bias, we can
rewrite Equation (\ref{eq:corr}) to obtain:

\be
\Delta \xi(\theta)= \xi_{\delta m} (\theta) \times \Big(1 +  
\frac{1}{P_{\gamma }^2}\frac{\xi_{+,{\rm PSF}}(\theta)}{\xi_+ (\theta)} \Big)\,,
\ee

\noindent where $\xi_{\delta m}(\theta)$ is the correlation function
of the residuals $\ave{\delta m (\thetag) \delta m (\thetag^\prime)
}$.  As for the uncorrected case, the first term is related to the
multiplicative bias, while the second term corresponds to the additive
bias.

\subsection{Bias estimation} 

We assume the calibration sample to be a contiguous patch with the same
area as the total area of the datsets presented in Table \ref{tab:2}.
This increases the sampling variance compared to what one would obtain
from several independent fields. Hence, our estimates are
conservative.  The model for the bias from this calibration subsample
is then used to correct the bias for the whole survey. To estimate the
sampling variance we generate 32 independent realisations where each
time a different location for the calibration sample is selected.

We create simulated shear catalogues for three different scenarios.
In the first scenario we do not correct for the bias, whereas in the
second case we use the calibration sample to derive an average bias
value as a function of the redshift. We sample the bias in steps of
$\Delta z=0.1$. In the third case we use the calibration sample to
determine the average bias as a function of redshift and observed
colour. We sample the bias in steps of $\Delta z=0.1$ and 15 bins of
colour of width 0.10 covering the range between $[0.05,1.45]$.

We start by exploring the $m$ and $c$ contributions to the bias in the
correlation function separately. The left panel of Figure
\ref{fig:m_and_c} shows the amplitude of $\xi_c(\theta)$ for the
tomographic bin $0.2<z<0.3$, which can be compared to the cosmic shear
signal $\xi_+(\theta)$ (black line). Since the PSF orientation is
constant, $\xi_c(\theta)$ is always positive, its asymptotic value is
$\ave{m}^2 e_{\rm PSF}^2$ and there is a scale for which the value of
$\xi_c$ becomes larger than the cosmological signal. This happens at
scales larger than $200 ~{\rm arcmin}$ even for the smallest redshift
bin. In reality $\xi_c(\theta)$ is expected to be a factor $\sim 10$
smaller as the {\it Euclid} PSF anisotropy in the currently agreed
optical design (c. 2011-2012) is supposed to be about $7\%$ only in
the corners while it is about $2-3\%$ in most of the field of view.
Additionally, the PSF pattern varies across the field, so that
$\xi_{\rm PSF}(\theta)$ decreases quickly and it is expected to be
very small on scales that are larger than the field of view (indicated
in the left panel of Figure \ref{fig:m_and_c} and in Figure
\ref{fig:bias_corr} by the magenta dashed lines).

As indicated by the green triangles in Figure~\ref{fig:m_and_c}, the
value for $\xi_c(\theta)$ is reduced by two orders of magnitude at
large angular scales when we use the average bias as a function of
redshift to correct the colour gradient bias. The improvement is
smaller on small scales, where the clustering of galaxies is
important. This is remedied by modelling the bias as a function of
colour and redshift (blue triangles). The error bars shown in
Figure~\ref{fig:m_and_c} have been computed from the 32 realisations
and include sampling variance.

The right panel of Figure \ref{fig:m_and_c} shows the value of $\Delta
\xi (\theta)$ when $\xi_c(\theta)=0$. As expected, when no correction
is applied, the overall bias $|\Delta \xi(\theta)|$ is about
$2|\ave{m}|$. Since $m<0$, the signal is suppressed (i.e $\Delta
\xi (\theta) <0 $). Once again, correcting the average bias as a
function of redshift improves the results. Including the colour
dependence improves the results even further to a level below the
requirement of $5\times 10^{-4}$ (indicated by the dotted line).
  
Figure~\ref{fig:bias_corr} shows $|\Delta \xi(\theta)|$, for three
redshift bins. If the colour gradient bias is ignored, $|\Delta
\xi (\theta)|$ has a minimum due to the competition between the
multiplicative and additive terms, which have an opposite effect on
the observed ellipticity correlation function. The location of the
minimum indicates the scale where the additive bias starts to dominate.
Similarly to what we found in Figure \ref{fig:m_and_c}, correcting the
measured shear by the average bias as a function of redshift reduces
the amplitude of the residual bias on the correlation function.
Taking into account the dependence upon the observed colour reduces
this residual further, in particular for small angular scales.  For
reference we indicate with a dotted line the maximum value for
$|\Delta \xi(\theta)|$ allocated to achieve {\it Euclid}'s science
objectives.

The measurements at different angular scales in
Figure~\ref{fig:bias_corr} are correlated, which complicates a simple
interpretation of the significance of the results. We therefore
computed the average residual bias, accounting for the correlation
between the angular scales using the covariance matrix computed from
the simulations. The results are reported in Table~\ref{tab:tabxi} for
the three redshift bins after correcting the bias either as a function
of redshift or as a function of both redshift and colour.  We derive
the average residual bias for three different ranges: the full range
of scales, i.e. $\theta < 240\, {\rm arcmin}$; eliminating scales
which will not be included in the cosmological analysis, i.e.  $\theta
> 6 \, {\rm arcmin}$; eliminating scales larger than the {\it Euclid}
field-of-view, i.e. $6 <\theta < 50\, {\rm arcmin}$. In all cases we
find that modelling the bias as a function of colour and redshift
produces better results than modelling the bias as a function of
redshift only. With this correction we obtain residual biases that are
consistent (within $1 \sigma$) with the maximum allowed value (i.e. $5
\times 10^{-4}$). A more realistic model for the PSF anisotropy would
result in a significantly smaller residual bias, especially on large
scales.

Overall these results demonstrate that the {\it HST} archive contains
a sufficient number of galaxies observed in at least two suitable
bands to calibrate the colour gradient-induced bias to the precision
required for {\it Euclid}. To simulate the spatial variation of the
bias using the Millennium Simulation, we assumed that the colour is
the main parameter to consider. We thus ignored the fact that the bias
also depends on other parameters such as its ellipticity, Sers\'ic
index, etc., as is evident from Figure \ref{fig:parameters}. This has
the effect of reducing the simulated scatter in the bias for a given
colour and redshift, but we accounted for this by including an
intrinsic scatter in order to match results from a preliminary
analysis of EGS data (Huang et al., in prep.). We note that we assumed a rather pessimistic PSF anisotropy and PSF size and therefore are
confident that our estimates are in fact conservative.  On real data,  we can can test further bias dependencies  and model it as a function of  an optimal set of  parameters.  The efficiency of the correction  can  be tested and eventually improved by dividing the {\it Euclid} dataset  in subsamples and comparing the differences in shear distributions measured on these subsamples.
 
\section{Conclusions}\label{sec:conclusions}

{\it Euclid} will measure the cosmic shear signal with unprecedented
precision \citep{redbook}. To do so it will measure galaxy shapes from
images observed in a rather broad filter. In combination with the
wavelength dependence of the PSF this leads to biases in the shape
measurements because the colours of galaxies vary spatially. To ensure that
the constraints on cosmological parameters are not compromised it is
important to quantify this `colour gradient' bias and examine possible
approaches to mitigate the problem.

In this paper we show how the spatial variation of the colour of a
galaxy generally leads to a bias in the measurement of its shape. In
the case of unweighted moments the PSF can be removed without
introducing any bias. In practice, in order to suppress the noise,
shear measurement methods use a weight function that assigns more
weight to the inner regions.  Consequently, the bias depends on the
choice of the weight function and is therefore generally
method-dependent. We derive results for moment-based methods, but note
that our findings are also applicable to methods that fit galaxy
models to the data.

We showed analytically that the amplitude of the bias scales
proportional to the square of the width of the filter used for the
measurement. We studied in detail how the bias depends on the
characteristics of the galaxies and the PSF using model galaxies.  The
amplitude of the bias depends strongly on the ratio of the observed
FWHM of the galaxy to the FWHM of the PSF, with a larger PSF leading
to a larger bias. The bias can be as large as a few percent if the
observed galaxy is about the size of the PSF, while it decreases
rapidly to only a few times $10^{-4}$ when the galaxy size is 1.6
times the PSF size. Given the small size of the {\it Euclid} PSF and
the magnitude limit of $24.5$, the observed galaxies will generally
have an observed FWHM which is larger than the PSF. The bias depends
on the bulge and disk sizes, on the ellipticity, but it depends most
strongly on the colour of the galaxy.

It is possible to determine the bias observationally using spatially
resolved data observed in at least two filters. We focus on the
particular case of {\it HST} observations in F606W and F814W, which
are best suited for this purpose. We quantify the limitations in
modelling the local SED using only two filters taking into account the {\it
  HST} PSF. We find that the linear interpolation we use to
approximate the local SED is the main source of error. This could be
improved in principle by performing a local SED fit.  This might be
possible especially for low redshift galaxies where we have resolved
observations in various colours. Errors in the model for the {\it HST}
PSF have a very minor impact. Our results therefore indicate that we
can model the bias due to colour gradients using {\it HST}
observations with an accuracy of a few times $10^{-4}$ for a typical
galaxy used in the {\it Euclid} weak lensing analysis.

To determine the average bias with sufficient precision requires a
representative sample of galaxies. \citet{Voigtetal12} suggested that
a sample of about thousand galaxies per redshift bin would be
sufficient to determine the bias with a precision of $\sigma_{\ave{m}}=
2 \times 10^{-4}$. This has been confirmed by a preliminary analysis
using EGS data (Huang et al. in prep.). The amount of archival {\it
  HST} data observed in at least two bands is sufficient for this.

A complication arises from the fact that the bias depends strongly on
the colour. As a result the clustering of galaxies will lead to
spatial variations in the bias. In this case an average correction for
each redshift bin may still lead to biases in the two-point
ellipticity correlation function. To examine the impact of this second
order effect we used a ray-tracing catalogue based on the Millennium
Simulation. We find that correcting the bias as a function of both
redshift and colour using a contiguous patch containing about 40,000
galaxies is sufficient to correct the bias to the accuracy required
for the {\it Euclid} two-point shear tomography.  This is a
conservative estimate of the number of galaxies in the {\it HST}
archive at the time of {\it Euclid}'s launch. We therefore conclude
that the presence of colour gradients in galaxies is not a limiting
factor for the {\it Euclid} cosmic shear analysis.

\section*{Acknowledgements}
We would like to thank Gary Bernstein, Christopher Hirata, J\'er\^ome
Amiaux and Frederic Courbin  and in general, the members of the {\it Euclid} Consortium  for helpful discussions.
  ES and HH
acknowledge the support of the Netherlands Organization for Scientific
Research (NWO) through a VIDI grant and the support from the European
Research Council under FP7 grant number 279396.  BJ acknowledges
support by a UK Space Agency Euclid grant and by STFC Consolidated
Grant ST/J001422/1.  TK was supported by a Royal Society University
Research Fellowship.  JR was supported by JPL, run by Caltech under a
contract for NASA.  TS acknowledges support from NSF through grant
AST-0444059-001, SAO through grant GO0-11147A, and DLR through grant
FKZ 50 QE 1103. MV acknowledges support from the Netherlands Organization for Scientific Research (NWO) and from the Beecroft Institute for Particle Astrophysics and Cosmology.
This material is based upon work supported in
part by the National Science Foundation under Grant No. 1066293 and
the hospitality of the Aspen Center for Physics.
\bibliographystyle{mn2e}
\bibliography{mybib}	 
%\bsp
\end{document}